\documentclass[12pt,twocolumn,preprint,a4]{iopart}
\usepackage[pdftex]{graphicx}
\usepackage{xcolor}
\usepackage[utf8]{inputenc}

\usepackage{siunitx}
\expandafter\let\csname equation*\endcsname\relax 
\expandafter\let\csname endequation*\endcsname\relax 
\usepackage{amsmath}
\newcommand{\Ha}{H$_{\alpha}$}
\newcommand{\Hb}{H$_{\beta}$}
\newcommand{\Hg}{H$_{\gamma}$}

\begin{document}

\title[]{
Ignition and propagation of nanosecond pulsed discharges in distilled water - negative vs. positive polarity applied to a pin electrode}

\author{K. Grosse, M. Falke, A. von Keudell}

\address{Experimental Physics II - Reactive Plasmas, Ruhr-Universit\"at Bochum, D-44780 Bochum, Germany}

\ead{Katharina.Grosse@rub.de}

\date{\today}

\begin{abstract}
Nanosecond plasmas in liquids are being used for water treatment, electrolysis or biomedical applications. The exact nature of these very dynamic plasmas and most important their ignition physics are strongly debated. The ignition itself may be explained by two competing hypothesis: (i) ignition in water may occur via field effects at the tip of the electrode followed by tunneling of electrons in between water molecules causing field ionization or (ii) via gaseous processes of electron multiplication in nanovoids that are created from liquid ruptures due to the strong electric field gradients. Both hypothesis are supported by theory, but experimental data are very sparse due to the difficulty to monitor the very fast processes in space and time. In this paper, we are using fast camera measurements and fast emission spectroscopy of nanosecond plasmas in water applying a positive and a negative polarity to a sharp tungsten electrode. It is shown that plasma ignition is dominated by field effects at the electrode-liquid interface either as field ionization for positive polarity or as field emission for negative polarity. This leads to a hot tungsten surface at a temperature of 7000 K for positive polarity, whereas the surface temperature is much lower for the negative polarity. At ignition, the electron density reaches 4 $\cdot$ 10$^{25}$ m$^{-3}$ for positive and only 2 $\cdot$ 10$^{25}$ m$^{-3}$ for the negative polarity. At the same time, the emission of the \Ha~light for the positive polarity is 4 times higher than that for the negative polarity. During plasma propagation, the electron densities are almost identical of the order of a 1 to 2 $\cdot$ 10$^{25}$ m$^{-3}$ and decay after the end of the pulse over 15 ns. It is concluded that plasma propagation is governed by field effects in a low density region that is created either by nanovoids or by density fluctuation in super critical water surrounding the electrode that is created by the pressure at the moment of plasma ignition.
\end{abstract}

\maketitle

\section{Introduction}

Plasmas in liquids is a prominent field of research being at the core of many applications ranging from wastewater treatment, to plasma supported electrolysis, to applications in medicine \cite{Bruggeman.2009}. The discharges are usually ignited by applying a high voltage (HV) pulse to a pin electrode immersed in the liquid opposite to a grounded counter electrode (either a pin or a plane electrode). The high voltage can be applied continuously or in the form of short pulses with a specific temporal structure, which strongly affects the ignition and plasma physics of these discharges. For example, if the pulse rise time is slow in the range of microseconds or longer, Joule heating of the liquid occurs at first and the local evaporation of the liquid will form small gas filled bubbles. The condition for ignition can then be reached in the gaseous environment and a plasma is formed that causes these bubbles to further expand and to initiate various chemical gas phase reactions. If, however, the pulse rise time is of the order of nanoseconds only, the inertia of the liquid adjacent to the pin electrode is so high, that any formation of a macroscopic bubble and any consecutive gas phase ignition of a plasma is not possible. Nevertheless, luminous discharges are being observed even on nanosecond time scales \cite{Seepersad.2013}. We refer to these discharge for simplicity as \textit{streamers}, although the physics behind might differ from traditional streamer discharges in gases. The understanding of the creation of such plasmas \textit{inside} the liquid is still an open question until today \cite{Joshi.2009, Zhuang.2016}, which is addressed by several ignition models: (i) in a seminal work, Seepersad et al. \cite{Seepersad.2013} postulated the presence of regions of low density in the liquid in front of a high voltage electrode that are induced by liquid ruptures caused by the high electric field gradients. Such ruptures induce the formation of so-called \textit{nanovoids}. It is striking that ignition only occurs above electric field strengths that allow also liquid rupture. The ignition itself may then proceed by a Townsend mechanism for charge multiplication inside these nanovoids. (ii) Plasma ignition inside the liquid may also occur in previously formed bubbles, which are present in the liquid due to an insufficient outgassing. When these microscopic gas bubbles exist close to the tip of the biased electrode, plasma ignition in these bubbles may occur \cite{Sharbaugh.1978}. Simek et al. \cite{Simek.2020, Hoffer.2020} used laser interferometry to analyse the plasma generation on a picosecond scale and observed very dense streamer shadows emerging from the electrode tip only for the very first pulse of an experiment that started with de-ionized and outgassed water. Even the second plasma pulse in the very same experiment showed a more inhomogeneous streamer pattern, suggesting that some of the created gas from the first plasma pulse remain and influences the ignition of all consecutive pulses. (iii) Alternatively, the classical breakdown in solids has been employed by Devins et al. \cite{Devins.1981} to describe plasma generation inside transformer oils using the Zener theory. Here, electron tunneling between adjacent liquid molecules acts as the source of ionization during plasma propagation \cite{OSullivan.2008}.

In a recent paper series \cite{Grosse.2020,vonKeudell.2020}, we analyzed the ignition phase of nanosecond plasmas in water for positive polarity applied to a tungsten wire electrode. The spectra are predominantly composed of a black body background originating from the emission of the hot tungsten electrode during the pulse. This tungsten electrode exhibits a temperature close to the boiling temperature of tungsten of around 7000 K depending on the pressure in front of the electrode. On top of this background, the Balmer series of hydrogen atoms becomes visible as well as the atomic oxygen line at 777 nm at later times after ignition. The modeling of the Balmer series indicated a strong contribution of self absorption, which has been modeled by two contributions to the emission pattern, one originating from the ionization zone at the streamer head with an extension in the range of a few micrometers only showing lines with weak self absorption line profiles and one originating from a recombination region in the streamer channel behind the head with an extension of 50 to 100 micrometers showing lines with strong self absorption line profiles. The observed degree of self absorption of the emission line profiles could only be explained assuming a very high density environment being close to the liquid density. Therefore, the hypothesis of field effects being also relevant for ignition and plasma propagation has been stated. The electron densities had been extracted from Stark broadening of the H$_{\alpha}$ lines yielding values up to a few 10$^{25}$ m$^{-3}$.

The interpretation of the observed strong self absorption of the emission lines as an indication of field effects governing plasma propagation is, however, only a very indirect conclusion. Therefore, we attempt in this paper to compare plasma propagation for different polarities of the voltages applied to the electrode immersed in water, because it is expected to affect the plasma characteristics. The most prominent example are the properties of positive versus negative streamer discharges in gases. For example, if ignition processes in the gas phase dominate for these nanosecond plasmas, those differences should also show up in the analysis of our experiments. For example, positive streamers exhibit a more diffuse structure in comparison to negative streamers since the electrons created by photo ionization in front of the streamer head are accelerated into it, whereas negative streamers are usually more localized, since the strong electric field creates a local ionization front \cite{Nijdam.2020,Fridman.2005}. In addition, the electron density in positive streamer heads is much lower compared to negative streamer heads.

The transfer of this knowledge regarding streamers in gases onto the case of plasma ignition at the electrode-liquid interface is not straightforward since two aspects need to be considered: (i) the temporal development of the medium going through the phases liquid-gas-plasma in correlation to the temporal development of the applied voltage, and (ii) the contribution of field emission of seed electrons directly at the tip of the electrodes. Four different models A, B, C and D for the dynamic of the plasma can be distinguished (see illustration in Fig. \ref{fig:threecases}):

\begin{figure}[ht]
    \centering
    	\includegraphics[width=8cm]{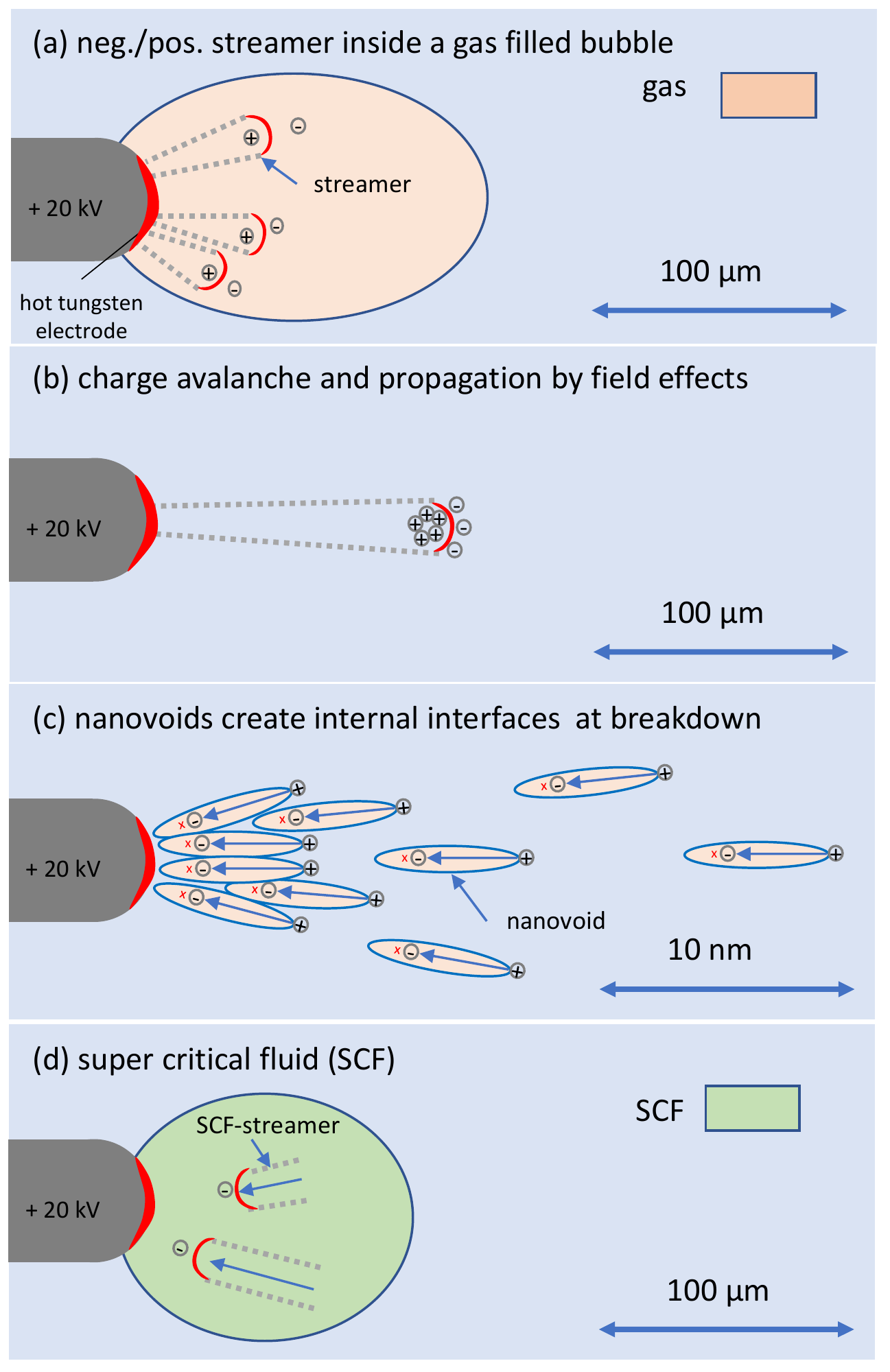}
    \caption{Models for plasma ignition and propagation as gaseous streamers inside gas filled bubbles in the liquid (model A), as a charge avalanche that propagates via field effects (model B), as charge multiplication inside nanovoids or at the interfaces between nanovoids and liquid (model C), or as electron channeling in a super critical fluid (SCF) (model D).}
    \label{fig:threecases}
\end{figure}

\begin{itemize}
\item \textit{Model A - Ignition via build up of gaseous streamers inside large gas filled bubbles inside the liquid (Fig. \ref{fig:threecases}(a))}: gas filled bubbles may be present from dissolved air in the liquid or may be created by Ohmic heating and evaporation during the rising time of the high voltage pulse.  When such gas filled bubbles are present, the development of electron avalanches in the gas phase is the most likely ignition mechanism. The build up of an avalanche and the formation of a streamer occurs over length scales of several 10s of micrometers to millimeters depending on the pressure. Tereshonok et al. \cite{Tereshonok.2017,Tereshonok.2018} analyzed the time span for the avalanche formation in conjunction to the formation of a gaseous bubble. Due to the finite time span for such a phase transition to occur, usually rather slow rising times (of the order of microseconds) and rather long pules are required for such mechanism to be prevalent. 

Different polarities of the voltage are expected to have a pronounced effect on the plasma properties similar to the difference between negative and positive corona discharges in gases and the associated differences between negative and positive streamers.

\item \textit{Model B - Ignition via charge avalanches that propagate via field effects (Fig. \ref{fig:threecases}(b))}: the formation of a gas filled bubble may be prohibited due the inertia of the liquid at very fast rise times of the voltage.  It is reasonable to assume that ionization and propagation of the discharge occurs via field emission or field ionization due to the strong gradient of the electric field that propagates through the material. Such a mechanism requires a very high electric field and is well known for semiconductor electron avalanche diodes. Field ionization of water molecules requires electric fields of at least 0.2 V/\AA\,\cite{Schmidt.1964, Anway.1969} that may occur in the vicinity of small protrusions at the electrode surface, but also at irregularities at the front of the propagating charge avalanche.

Different polarities of the voltage are expected to have no effect on \textit{plasma propagation} since the field effects are very local and the tunneling of an electron between adjacent water molecules should not depend on the direction of plasma propagation in first order. However, the different polarities might have an effect on \textit{plasma ignition} at the electrode liquid interface due to the difference between field ionization for a positive polarity versus field emission for a negative polarity applied to the electrode.

\item \textit{Model C - Ignition via charge multiplication inside nanovoids or at the interfaces between nanovoids and water (Fig. \ref{fig:threecases}(c)):} the high electric fields that are formed by applying a voltage to the sharp electrodes also causes ruptures in the liquid \cite{Shneider.2012, Shneider.2013}. This creates nanometer sized voids in the medium that may favor ignition and plasma propagation. The size of these nanovoids, that have not yet been observed so far, are expected to be smaller than required for the build up of a charge multiplication avalanche. But, field emission of electrons at the internal surfaces of the nanovoids, the acceleration of the electrons in the free space of these nanovoids and ionization at the opposite end may contribute to the ionization efficiency. Several models tried to analyse such a plasma propagation mechanism \cite{Li.2020,Aghdam.2020}. Any experimental verification is still lacking. 

Different polarities of the voltage may have an effect since the electrons are either accelerated inside the nanovoids towards the electrode or in the opposite direction. If the shape of these nanovoids is affected by the in-homogeneous electric field gradient, any differences comparing both polarities could become visible.  

\item \textit{Model D - Ignition via charge multiplication in super critical fluid (SCF) of water (Fig. \ref{fig:threecases}d):} the existence of regions of lower density might also be explained by the fact that water is expected to be in the super critical state due to the very high pressures upon plasma ignition. This state is also referred to as a super critical fluid (SCF) or as a \textit{cluster fluid} \cite{Stauss.2018}, since the molecules are not homogeneously distributed, but form rather clusters with space in between. The critical temperature of water is 647 K and the critical pressure is 2.2 $\cdot$ 10$^7$ Pa. The pressure at the moment of ignition reaches value up to 10$^9$ Pa \cite{Grosse.2019} in our experiment, which is well above that threshold. It can be assumed that plasma ignition occurs at first by field effects at the electrode-liquid interface leading to such high pressures and temperatures at the electrode tip initially. These pressures and temperatures are the boundary condition for a spherical pressure and temperature field that propagates away from the tungsten tip. Since pressure and temperature are expected to decrease with distance, one can conclude that also a spherical region of water at exactly the critical point has to be present. At its critical point, the medium exhibits strong density fluctuations which may support ionization and acceleration of electrons in the high electric fields via electron channeling \cite{Muneoka.2015}. This has been studied for the breakdown at the critical point for CO$_2$, where a reduction of the ignition voltage at the critical point was observed \cite{Ito.2002,Muneoka.2015,Stauss.2018}. The study of plasmas in SCFs is usually performed by transferring the medium at first in the super critical state in a pressurized cell and then igniting a plasma. In the context of nanosecond plasmas in liquids, it is conceivable to assume that the ignition process creates this high pressure region initially in which then the plasma may propagate. This should be a self amplified process, since any local plasma creation will instantly bring a neighboring region of water in its super critical state.   

Different polarities of the voltage may have an effect since the electrons are either accelerated inside the SCF away from the electrode tip or towards the electrode tip. Since the charge multiplication is a very local process the electron densities are expected to be the same. But  the amount of excited volume might be different, as the electron avalanches propagate towards each other for positive polarity (see schematic in Fig. \ref{fig:threecases}d) and propagate apart for negative polarity.
\end{itemize}

Given the short time scale of the nanosecond pulsed plasmas, it is reasonable to assume that the ignition physics should properly described by models B, C, or D in Fig. \ref{fig:threecases}. Nevertheless, different electron densities are expected for the same voltage that is being applied to the electrode tip. For example in model A, the formation of positive streamers for a positive voltage applied to the electrode should lead to a smaller electron density compared to the case of negative polarity. In model B, the electron density should not be affected by the polarity of the voltage. In model C the electron density for the positive polarity might be higher, because the electrons are now accelerated into regions of larger nanovoid density compared to the negative polarity. In model D, the electron densities are expected to be identical. The validation of model C is most difficult and field effects at the internal interfaces of the nanovoids are overlapping with ionization due to the free acceleration of electrons inside the nanovoids. Both effects may counteract or amplify each other.

The distinction of models A, B, C, and D has to be regarded with great care since the dominance of each of the models may change with time during the pulses. For example, at the very beginning of the pulse on the time scale of picoseconds,  field effects should dominate leading to a description by model B. As the time progresses, liquid ruptures and/or the transition to an SCF may occur and the effects according to model C and D may contribute. Finally, for at least very long pulses, the coalescence of nanovoids and the expansion of the gas filled bubbles will eventually lead to a dominant contribution to the plasma formation according to model A. 

Next to the ionization and the plasma propagation mechanisms in the liquids/gas/plasma phase, also \textit{field effects} at the electrode surface change when comparing negative and positive polarity. In the case of negative polarity, the electrode tip is a continuous source of electrons, which are emitted due to the field effect at protrusions on the surface. This is very efficient and very high electron currents are being expected (similar to field emitter tips). This is in contrast to the positive polarity, where the field effect may ionize adjacent water molecules close to the surface at the onset of the high voltage pulse. However, any reduction in the density of the water molecules due the phase transition to the gas phase will drastically reduce the efficiency of charge transfer from the medium onto the electrode tip. 

Summarizing, one may conclude that a comparison between a negative and a positive polarity voltage applied to a pin electrode immersed in water based on an evaluation of the temporal development of the electron density and its absolute value may provide information regarding the dominance of one of the four models above. Four experimental quantities will be assessed in this paper: (i) the dissipated power is deduced from the difference between the forward and backward traveling HV pulses that are generated by a nanosecond HV pulser and sent along a transmission line to the plasma electrode, (ii) the total deposited energy depending on the polarity is evaluated using cavitation modeling of the bubble radius expansion versus time, (iii) the temporal development of the electron density is monitored from the Stark broadening of the H$_{\alpha}$ Balmer lines, and (iv) the degree of self-absorption of the hydrogen Balmer lines us used to assess the local density at the location of plasma emission. By comparing these quantities, the validity of models A, B, C, or D will be judged.

\newpage
\section{Experiment}

\subsection{Experimental setup}

\begin{figure}[ht]
    \centering
    	\includegraphics[width=11cm]{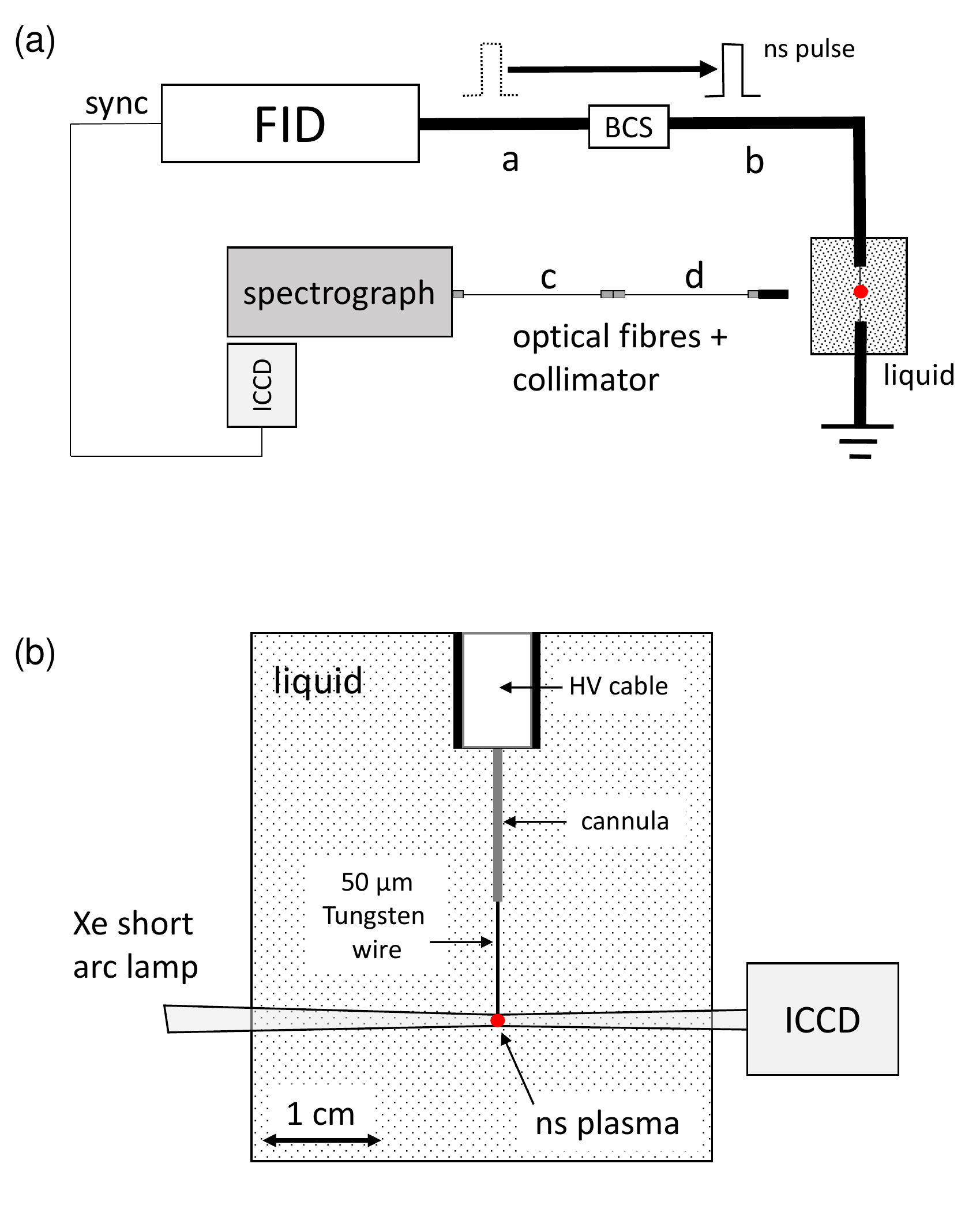}
    \caption{Sketch of the experimental setup. (a) Layout of the powering of the setup using a nanpsecond pulser from FID connected to the tungsten electrode with a HV cable where a back current shunt (BCS) is inserted. (b) Layout of the plasma itself used for shadowgraphy measurements.}
    \label{fig:setup}
\end{figure}

A detailed description of the setup is published elsewhere \cite{Grosse.2019} and the key elements are only shortly summarized in the following.
Fig. \ref{fig:setup} shows a schematic of the experimental setup used for time-resolved optical emission spectroscopy of the nanosecond pulsed plasma. The high voltage (HV) pulses are generated by two FID pulse generators (FID Technology GmbH) of different polarities, respectively. The pulses have a rising time of 2-3\,ns and a pulse width of 10\,ns. A frequency of 15\,Hz is applied as well as applied voltages of +20\,kV or -20\,kV, respectively. The cable connecting the power supply and the plasma electrode is 10\,m long with a back current shunt (BCS) mounted at a central position along the cable (with a\,=b\,=\,5m in Fig. \ref{fig:setup}a). The BCS consists of 11x3.3\,$\Omega$ resistors welded into the cable shield.

The plasma chamber is made of PMMA and three quartz windows are mounted, two from the sides for (shadowgraphic and ICCD) imaging and one at the front for the collection of light emission during optical emission spectroscopy. The electrodes are positioned at the center of the chamber approximately 10\,mm apart, as illustrated in Fig.\ref{fig:setup}b. Distilled water with an electrical conductivity of 1\,$\mu$ S cm$^{-1}$ and a pH of approximately 5.5 is used as the liquid. The plasma chamber and the FID pulser are both surrounded by a common Faraday cage so that electromagnetic interference (EMI) cannot escape the system.

\subsection{Power measurements}

The power dissipation in these nanosecond plasmas depends sensitively on the electric circuitry connecting the power supply to the plasma electrode. The electric pulses generated by the power supply travel along the cable to the electrode tip and generate the plasma but a part of the electric signal is also reflected, travels back to the power supply and is reflected again, before it re-ignites the plasma after a time span defined by the travel time along the cable. This yields an oscillation of the electrical power imposed onto the plasma with period times that can be comparable to the original length of the plasma pulse, as discussed in the literature \cite{Pongrac.2018,Marinov.2014,Grosse.2019}. This can be avoided by using either a matching of the impedance of the plasma by placing a resistor in parallel and in series to the tungsten electrode or by using a very long cable length so that the reflected pulse re-ignites the plasma at much later times compared to the first pulse. Here, we used a cable length of 10 m, which yields a delay in between these pulses of 100 ns, which is very much larger than the time span of the plasma development due to the ignition by the first pulse. 

The BCS measures the voltage of the traveling high voltage pulse generated by the pulser in the middle of the transmission line. When this pulse is reflected at the tungsten electrode, it is reflected and the forward traveling and the backward traveling pulses interfere due to the pulse reflection. This is taken into account by reconstructing the voltage at the electrode (in the following: \textit{electrode voltage}) from the BCS signals as follows: The initial and reflected voltage pulses measured at the BCS are summed up with respect to their temporal evolution. This estimation of the electrode voltage results in a peak amplitude of twice the applied voltage amplitude. 

The power dissipated in the discharge is assessed from the BCS data by multiplying the voltage and current measured at the BCS location according to the procedure by Simek et al. \cite{Simek.2017}. The absorbed power is then calculated by the difference between the power of the HV pulse before and after reflection at the electrode and plasma generation. 

\subsection{Shadowgraphy measurements and cavitation modeling}

Shadowgraphy images are taken after the discharge ignition to observe the gas bubble development in the liquid medium. Therefore, the electrode tip is illuminated by the light of a Xe short arc lamp and the light is focused onto the ICCD chip of an Andor iStar DH734-18U-03 camera with a lens system. The spectral sensitivity of the camera ranges from 180-850\,nm and was triggered by the sync-output of the HV pulsers. The jitter of only a few ps is negligible. The gas bubble expansion and collapse has been monitored by single-shot images with camera gates of 50\,ns .

The plasma induced bubble formation is modeled by cavitation theory, beginning with a bubble with radius $R$ and pressure $p$ at the interface between gas and liquid at time $t=0$. The pressure in the liquid far away from the bubble is $p_{\infty}$. Due to a varying pressure inside the liquid, the sound speed also varies with distance radius $r$. The expansion of a bubble can be modeled with the well known Rayleigh-Plesset equation \cite{GilmoreF.R..1952,Plesset.1949,Plesset.1977,Keller.1980,Keller.1956}:

\begin{equation}
R \ddot{R}\left(1-\frac{\dot{R}}{c}\right) + \frac{3}{2}  \dot{R}^2 \left(1-\frac{\dot{R}}{3c}\right) = h \left(1+\frac{\dot{R}}{c}\right) + \left(1-\frac{\dot{R}}{c}\right)\frac{R}{c}\frac{\partial h}{\partial t}\label{eq:rp}
\end{equation}

With $R$ the radius of the interface between bubble and liquid, $c$ the speed of sound at the location of that interface. $h$ denotes the enthalpy that is calculated from the equation of state for water as it depends on the pressure $p$. The experiments in our setup indicate that the velocity $c$ of the emerging acoustic waves depending on $p$ is significantly larger than the velocity $\dot{R}$ of the bubble radius itself. Therefore, we neglect the term proportional to $\frac{\partial h}{\partial t}$ in eq. \ref{eq:rp}. The pressure inside the bubble is given by the adiabatic expansion of the water vapor starting with an initial pressure $p_{0,gas}$ that constitutes a free parameter for the modeling. The initial radius is set to $R_0=$ 25 $\mu$m corresponding to a tungsten tip diameter of 50 $\mu$m. Equation \ref{eq:rp} is solved numerically with the boundary condition of $R_{t=0}=R_0$ and $\dot{R}_{t=0}=0$. All details of this calculation can be found in \cite{Grosse.2019}.

\subsection{Emission spectroscopy}

\subsubsection{Spectra acquisition and spectral features}~\\

 The emission spectra are acquired by an Andor iStar DH734-18U-03 camera that is mounted to a triple-grating SpectraPro 750 spectrograph from Acton Research. A 50 $\frac{grooves}{mm}$ grating blazed at 600 nm is used. The measurements were performed with a gate time of $t_{gate}=2\,ns$ and time steps of $t_{step}=2\,ns$ between the spectra. Therefore a time span of 50\,ns could be monitored from ignition to the end of the initial pulse. Due to the size of the CCD chip, the spectrum is composed of three different spectra with the central wavelengths (CW) of 400\,nm, 550\,nm and 850\,nm, covering a wavelength range of approx. 600\,nm. The spectra were background subtracted and calibrated with a broadband D-Halogen lamp. Due to internal reflections of the 2nd order onto the grating, a high pass filter was used for the measurements with CW of 550\,nm and 850\,nm. The data were processed with a MATLAB script and merged for every step in time. Each spectrum is averaged over 2000 discharges. The acquisition of spectra is performed using a glass fibre that collects light from the complete region including electrode tip and propagating streamer. Therefore, the light of all propagating plasmas in these 2000 discharges is properly collected despite the fact that the streamers propagate in individual discharges in random directions. 

\begin{figure}[ht]
    \centering
    \includegraphics[width=8cm]{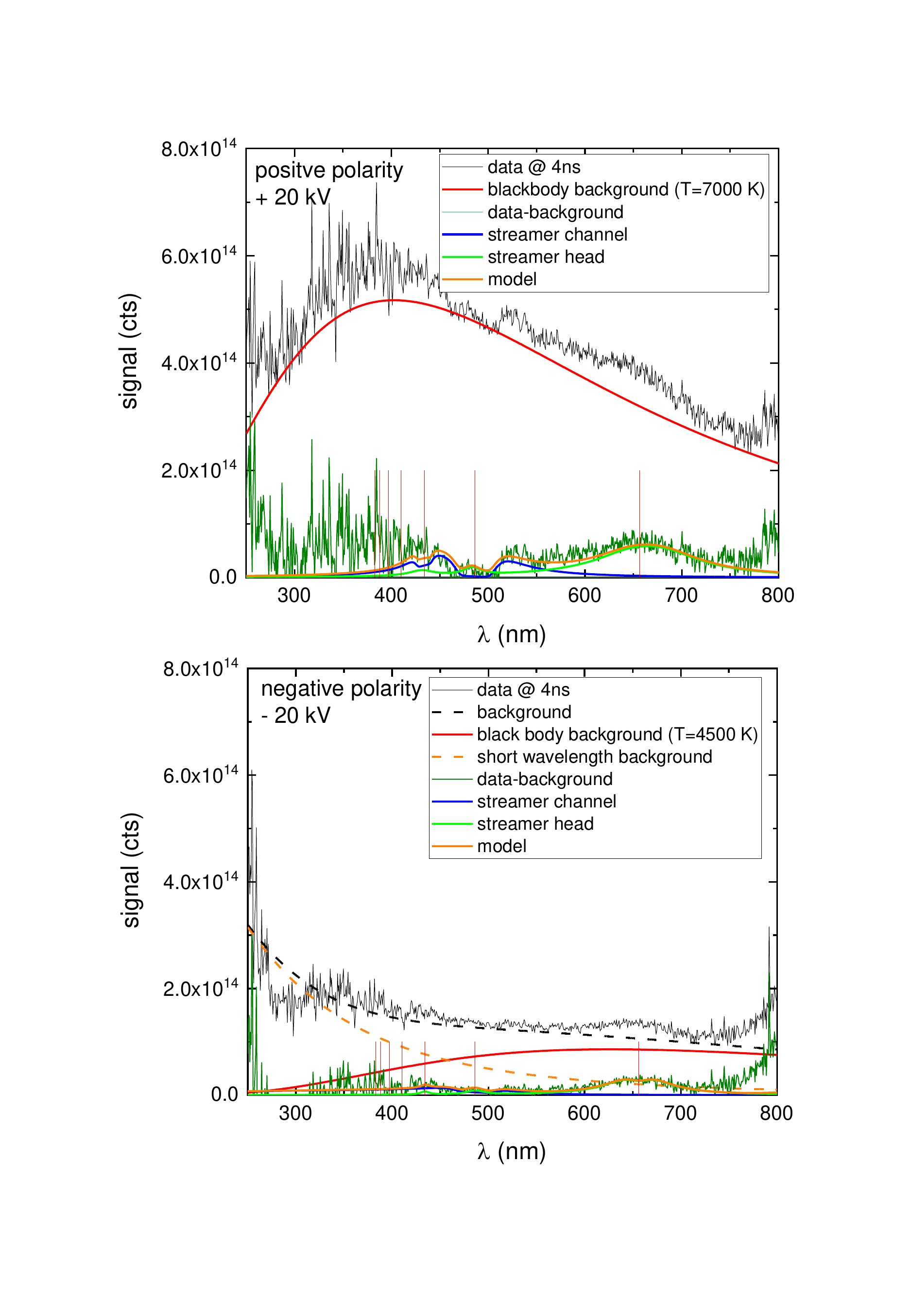}
    \caption{Example for analysis of an emission spectrum at 4ns after ignition for a polarity of +20 kV (a) and of -20 kV (b), respectively. The background is modeled by black body radiation assuming different temperatures, as indicated. The background corrected spectrum are modeled with line profiles of the Balmer series of hydrogen (vertical red thin bars) combining a contribution from the streamer channel and the streamer head. For details see text.\label{fig:spectrum2}}
\end{figure}

The emission spectra consist of a broad continuum and of broadened lines, as illustrated for an experiment monitoring the nanosecond plasmas at 4 ns after ignition for a polarity of + 20 kV and of -20 kV (generator output voltage) in Fig. \ref{fig:spectrum2}a and b, respectively. The spectra contain a contribution from black body radiation from the tungsten tip, which is also the brightest spot in the single shot images. This background may also contain a short wavelength contribution that may originate from any hot spots at the electrode (and thus being modeled also with black body radiation with a high temperature) or from a Bremsstrahlung contribution due to the acceleration of the electrons in collisions with ions and neutrals. The line emissions contribute only to a small percentage to the total number of generated photons. The broadened lines are from the hydrogen Balmer series (line positions indicated as thin red bars in Fig. \ref{fig:spectrum2}).

\subsubsection{Continuum radiation}~\\

The spectra are analysed and quantified following the procedure, as discussed previously in \cite{vonKeudell.2020,Grosse.2020}. The continuum contains a contribution of black body radiation of the hot tungsten tip, which can be clearly identified in the spectrum using a positive polarity in Fig. \ref{fig:spectrum2}a. The intense heating of tungsten can be linked to the electron current that is drawn by the positively biased electrode causing typical black body temperatures of 7000 K. The interpretation of the background as black body spectrum with a single temperature cannot easily be applied for the spectra collected for negative polarity, as illustrated in Fig. \ref{fig:spectrum2}b. Instead a short wavelength contribution is visible in all spectra. Such short wavelength contributions may originate from the formation of a hot spot on the electrode, as already discussed in \cite{Grosse.2020} for plasmas using a positive polarity at the electrode, but may also originate from Bremsstrahlung contribution due to the acceleration of the electrons in collisions with ions and neutrals. The limited signal-to-noise ratio in the data does not allow a clear distinction between black body radiation or Bremsstrahlung radiation in the short wavelength region. Furthermore, the signal calibration especially at short and long wavelengths below 280 nm and above 800 nm respectively suffers from the low detector sensitivity at these wavelengths. Therefore, the spectrum is modeled by assuming a short wavelength contribution (orange dashed line in Fig. \ref{fig:spectrum2}b) corresponding to black body at 20000 K and a long wavelength contribution corresponding to a varying temperature as indicated in Fig. \ref{fig:spectrum2}b. Both contributions are weighted to fit the background. This background fitting approach for the data from the experiments for negative polarity is somewhat arbitrary and the short wavelength contribution is not discussed any further.

\subsubsection{Hydrogen Balmer line emission}~\\

The line contribution to the spectra has been thoroughly analyzed in \cite{vonKeudell.2020}. Since the plasma is ignited directly in distilled water, line emissions of only H and O containing atoms and molecules are expected. Most prominent is the hydrogen Balmer series, as shown  
Fig. \ref{fig:spectrum2} for \Ha, \Hb~and \Hg~(Balmer series indicated as red bars). Due to the high temperatures and high densities of electrons and neutrals, all emission lines are affected by Doppler broadening, van der Waals broadening, or Stark broadening. Due to the very high electron densities, Stark broadening dominates, as already discussed in \cite{vonKeudell.2020}. This broadening effect is quantified by using line profiles calculated by Gigosos et al. \cite{Gigosos.2003} in the form of tables for FWHMs of the hydrogen Balmer series for different temperatures. The red shift of the lines due to the Stark effect is usually rather small and typically only 10\% of the FWHM. The example spectra in Fig. \ref{fig:spectrum2} are taken at 4 ns after ignition for which the plasmas show a peak in intensity. The broadened \Ha~lines are clearly visible, with the lines for the \Ha~lines for the positive polarity being a bit broader than for the negative polarity indicating a smaller electron density in the latter case.

The high density of species in the nanosecond plasma causes also self absorption of the emission lines  \cite{Cowan.1948}. This self absorption effect can be seen in the spectrum shown in Fig. \ref{fig:spectrum2}a, and especially in Fig. \ref{fig:spectrum2}b, as small line reversals of the \Ha~line. Self-absorption of an emission line takes  into account that the emitted photons of a transition may be reabsorbed by the very same species along the optical path. Self absorption by hydrogen atoms generated by the dissociation of water along the streamer channel may also absorb photons from the black body radiation of the hot tungsten tip. Due to the very local nature of the streamer propagation it is reasonable to assume that the likelihood that the streamer propagates along the line-of-sight of the optical path is very small. Therefore, blackbody radiation from the hot tungsten tip represents a non obstructed background in the spectra.

The Balmer \Ha, \Hb, and \Hg~lines are modeled by assuming different FWHMs and amplitudes, as well as a specific degree of self absorption. Two contribution are postulated, one originating from the ionization zone of the streamer showing very little self absorption (bright green line in Fig. \ref{fig:spectrum2}) and one originating from the streamer channel showing very large self absorption (blue line in Fig. \ref{fig:spectrum2}). The sum of those two contributions (orange line in Fig. \ref{fig:spectrum2}) is fitted to the background subtracted spectra (thin green line in Fig.  \ref{fig:spectrum2}). Details of this line fitting procedure are discussed in \cite{vonKeudell.2020}. Here, we analyse only the \Ha~line profiles and intensity that can easily be separated from the background continuum in the data. Thereby, this line analysis is also independent from any assumptions regarding the nature of the background continuum.

\newpage
\section{Results}

\subsection{Electrical characteristics and plasma emission}

\begin{figure}[ht]
    \centering
    	\includegraphics[width=11cm]{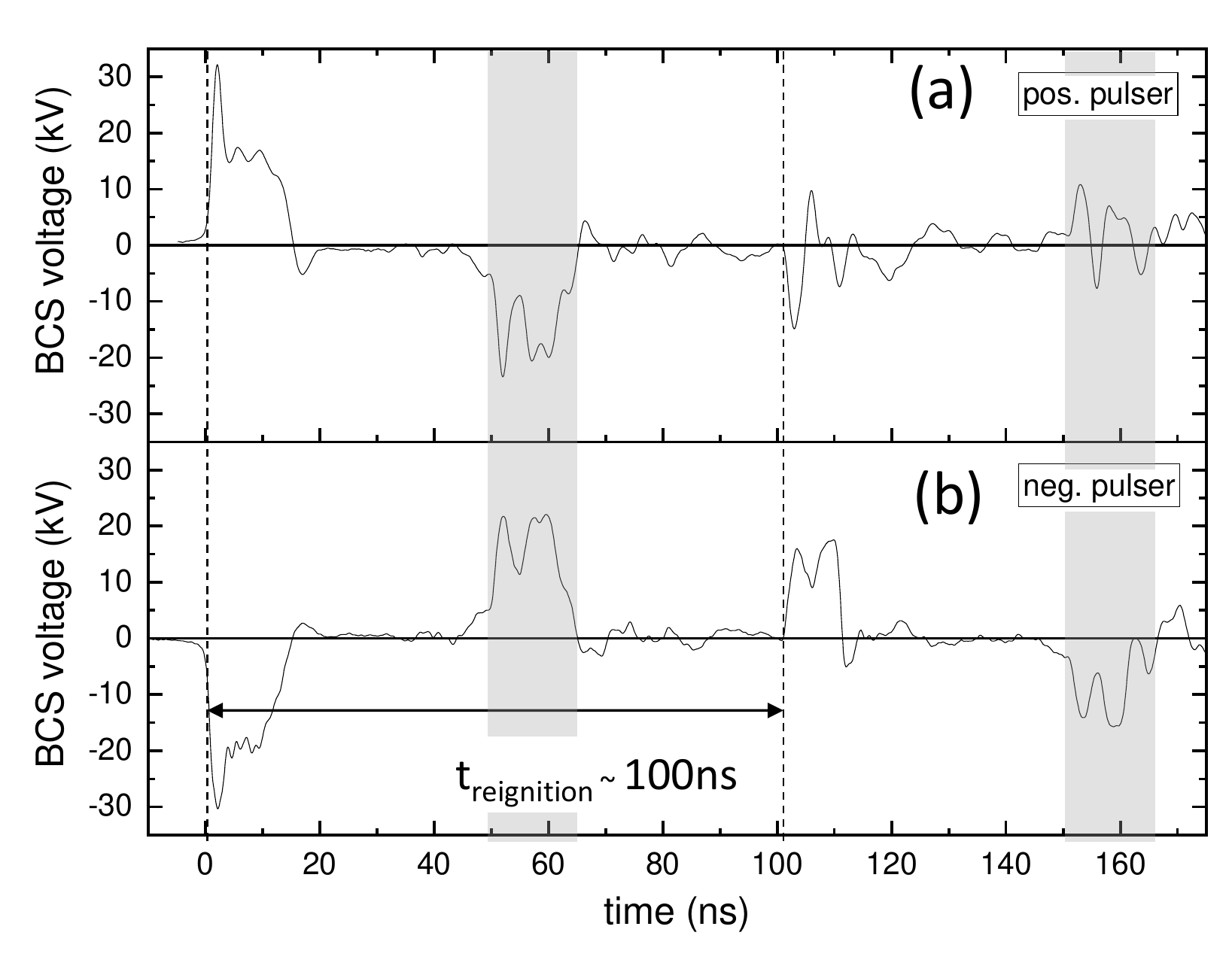}
    \caption{Voltages measured at BCS for positive (top) and negative (bottom) ns pulser with generator output voltage of 20\,kV. The gray areas mark the reflected pulses traveling back towards the pulser. The time between two forward traveling pulses leading to re-ignition of discharge is approximately 100\,ns.}
    \label{fig:voltagedata}
\end{figure}

The temporal evolution of the BCS voltage for the positive and negative pulser at $\pm$ 20\,kV generator output voltage is presented in Fig. \ref{fig:voltagedata}. The gray areas show the time window when the pulses that are reflected at the electrode tip are measured again at the BCS location. The reflected pulses for both polarities show two distinct peaks, whereas the forward traveling pulses show only one peak at the very beginning. The consecutive reflections at the pulser change the signals and the pulses re-appear at the BCS location after 100\,ns (indicated by dashed line in Fig. \ref{fig:voltagedata}). The decrease in voltage when comparing the forward with the reflected pulses is used to quantify the dissipated energy, yielding 12 mJ for the positive and 18 mJ for the negative polarity. 

\begin{figure}[ht]
    \centering
	\includegraphics[width=15cm]{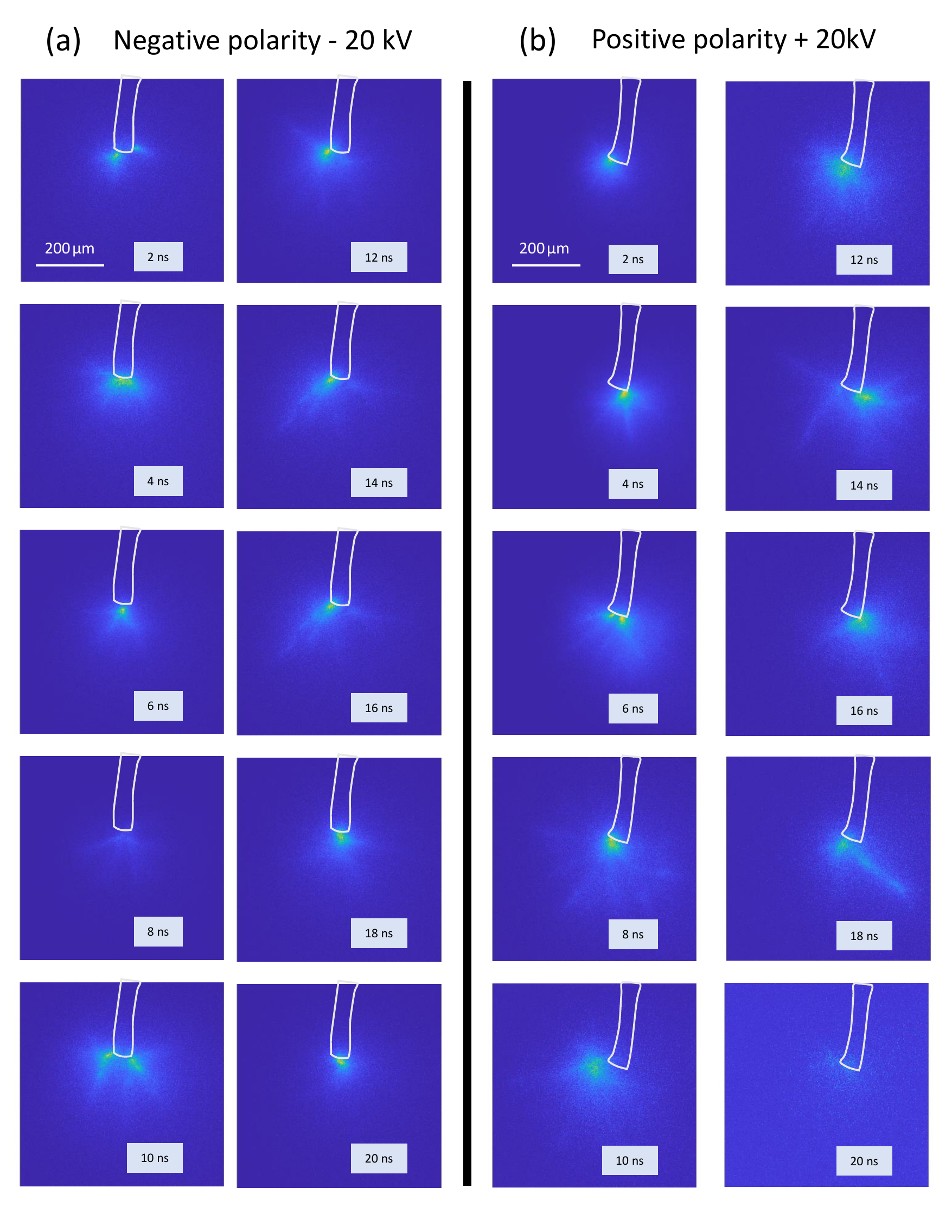}
    \caption{Images of plasma emission for negative (left) and positive polarity (right). The individual times are indicated. The white line denotes the shape of the electrode. Each image is scaled to its maximum intensity for best visibility.}
    \label{fig:lightemission}
\end{figure}

The temporal evolution of the voltage is also compared to ICCD images of plasma emission for negative and positive polarity of -20 kV and of +20 kV in Fig. \ref{fig:lightemission}. The images are taken during different pulses with an adjusted and shifted time delay. It can be seen that several ignition locations at the electrode tip can become visible for both polarities that also seem to illuminate several streamer channels propagating away from the electrode. The plasma pattern in the ICCD images appear for later stages of plasma propagation rather similar for both polarities. The location of the electrode tip itself remains the brightest spot of emission independent of the polarity of the voltage. This is consistent with the observation of a dominant contribution from black body radiation in the spectra (see example spectra in Fig. \ref{fig:spectrum2}a) that originates from the hot tungsten surface, as already discussed in \cite{Grosse.2019}.  

\begin{figure}[ht]
    \centering
	\includegraphics[width=11cm]{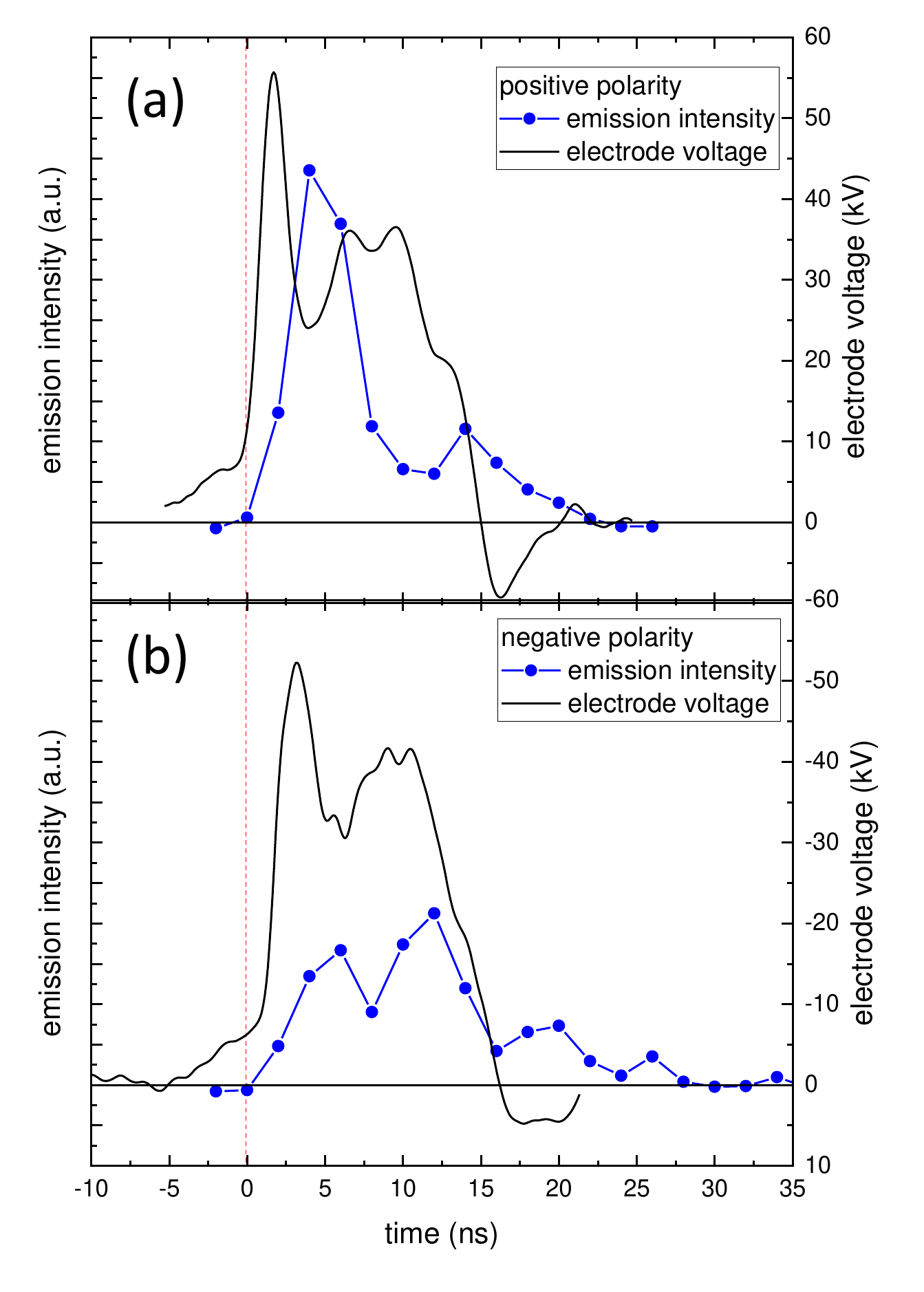}
    \caption{Reconstructed voltage pulse from BCS data (black line) for 20\,kV generator output voltage for the positive (a) and negative (b) pulser. The integrated light intensity is shown in blue. The red vertical line denotes $t$ = 0 ns}
    \label{fig:voltageemission}
\end{figure}

The electrode voltage is reconstructed from the interference of the forward and backward traveling HV pulse upon reflection at the electrode. The integrated emission measured in the ICCD of Fig. \ref{fig:lightemission} is then correlated to the shape of the electrode voltage, as presented in Fig. \ref{fig:voltageemission}.

One can clearly see two emission maxima at the rising and falling edge of the positive and the negative high voltage pulse. In between, a dark phase of the pulse, as already reported in the literature \cite{Seepersad.2013}, can be identified. 

Although the electrode voltages look similar, the temporal evolution of the emission intensity differs when comparing both polarities. The first emission intensity maximum using the positive pulse is the strongest, whereas the second intensity maximum is higher for the negative pulse. This could indicate, that plasma ignition during the rise of the voltage pulses depends on the polarity. 

\subsection{Bubble formation depending on electrode polarity}

\begin{figure}[ht]
    \centering
    \includegraphics[width=11cm]{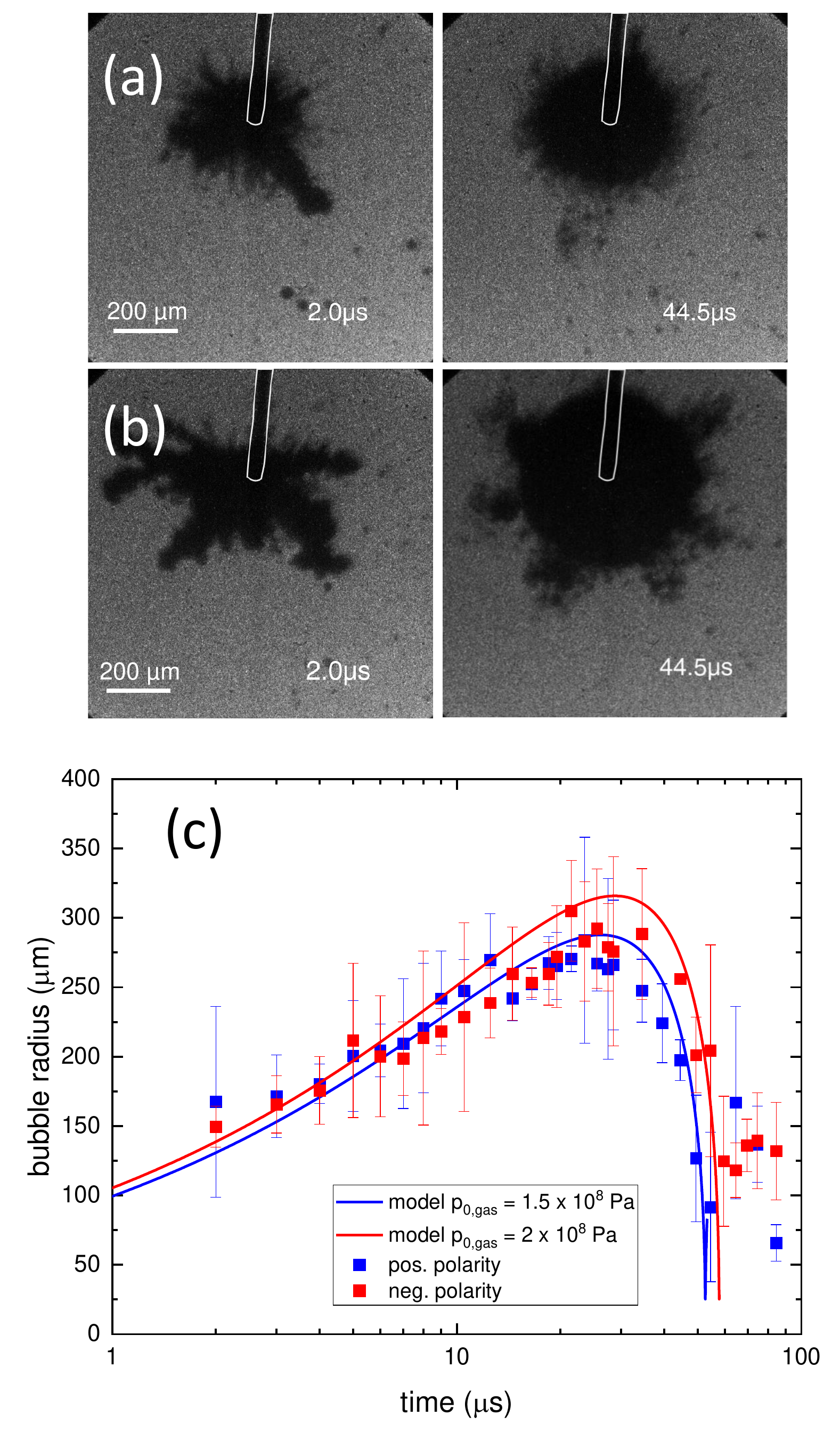}
    \caption{Shadowgraphy images for the negative (a) and positive (b) polarity at different times after plasma ignition. The white line marks the shadow of the electrode. (c) Bubble radius vs. time for negative (black squares) and positive (red squares) polarity. The solid line denotes a Rayleigh-Plesset modelling of bubble expansion using an initial pressure of 1 GPa and a volume energy of 1 $\cdot$ 10$^{-5}$ J (blue line) and of 1.3 $\cdot$ 10$^{-5}$ J (red line).}
    \label{fig:cavitation}
\end{figure}

The nanosecond plasma is eventually initiating the expansion of a cavitation bubble on time scales of microseconds. Selected shadowgraphs are shown in Fig. \ref{fig:cavitation} that are taken at different times after ignition for negative polarity of -20 kV (a) and for positive polarity of +20 kV (b). Fig. \ref{fig:cavitation}c shows the temporal evolution of the bubble radius for both polarities in comparison with cavitation modeling using an initial pressure $p_{0,gas}$ of 1.5 $\cdot$ 10$^8$ Pa and of 2 $\cdot$ 10$^8$ Pa for an initial volume with radius $R_0$ of 25 $\cdot$ 10$^{-6}$ m. The pressure in the ambient liquid has been set to $p_{\infty}$ = 10$^5$ Pa. Details are discussed in \cite{Grosse.2019}.

The dissipated energy given by the product of initial pressure $p_{0,gas}$ of 2 $\cdot$ 10$^8$ Pa times volume of a sphere with radius $R_0$ yields 1.3 $\cdot$ 10$^{-5}$ J, which is much smaller than the dissipated energy of the order of 10$^{-2}$ J that has been derived from comparing the BCS data of the traveling high voltage pulses. Such a difference is not surprising, since the initial pressure $p_{0,gas}$ refers to the kinetic energy of species that are created by water dissociation in the plasma. The energy that is dissipated into dissociation, into heating the tungsten surface, or into photon emission does not instantly contribute to the pressure that drives the expansion of the cavitation bubble. 

The comparison between the cavitation modeling and the observed bubble radius shows some deviations especially at times earlier than 10 $\mu$s and at times later than 50 $\mu$s: (i) at earlier times, any differences may be associated with the fact that the initial expansion follows the streamer channels that are not spherical (as shown in Fig. \ref{fig:cavitation}a and b) as the spherical bubble in the model; (ii) at later times, when the bubble collapses again, the decrease in bubble sizes is much faster in the model compared to the data. This may be associated with the fact that the collapse of the bubble occurs around a tungsten wire, whereas the model may be too simple for this phase since it regards only a collapsing spherical bubble without any adjacent solid surface. 

It is, however, most striking that the temporal evolution of the bubble radius does not depend on the polarity of the voltage applied to the electrode. Apparently, the energy that is dissipated in the streamers propagating from the electrode into the liquid that eventually convert water into a hot gas that drives cavity expansion is the same.

\subsection{Microstructure of the tungsten electrodes}

\begin{figure}[ht]
    \centering
    	\includegraphics[width=11cm]{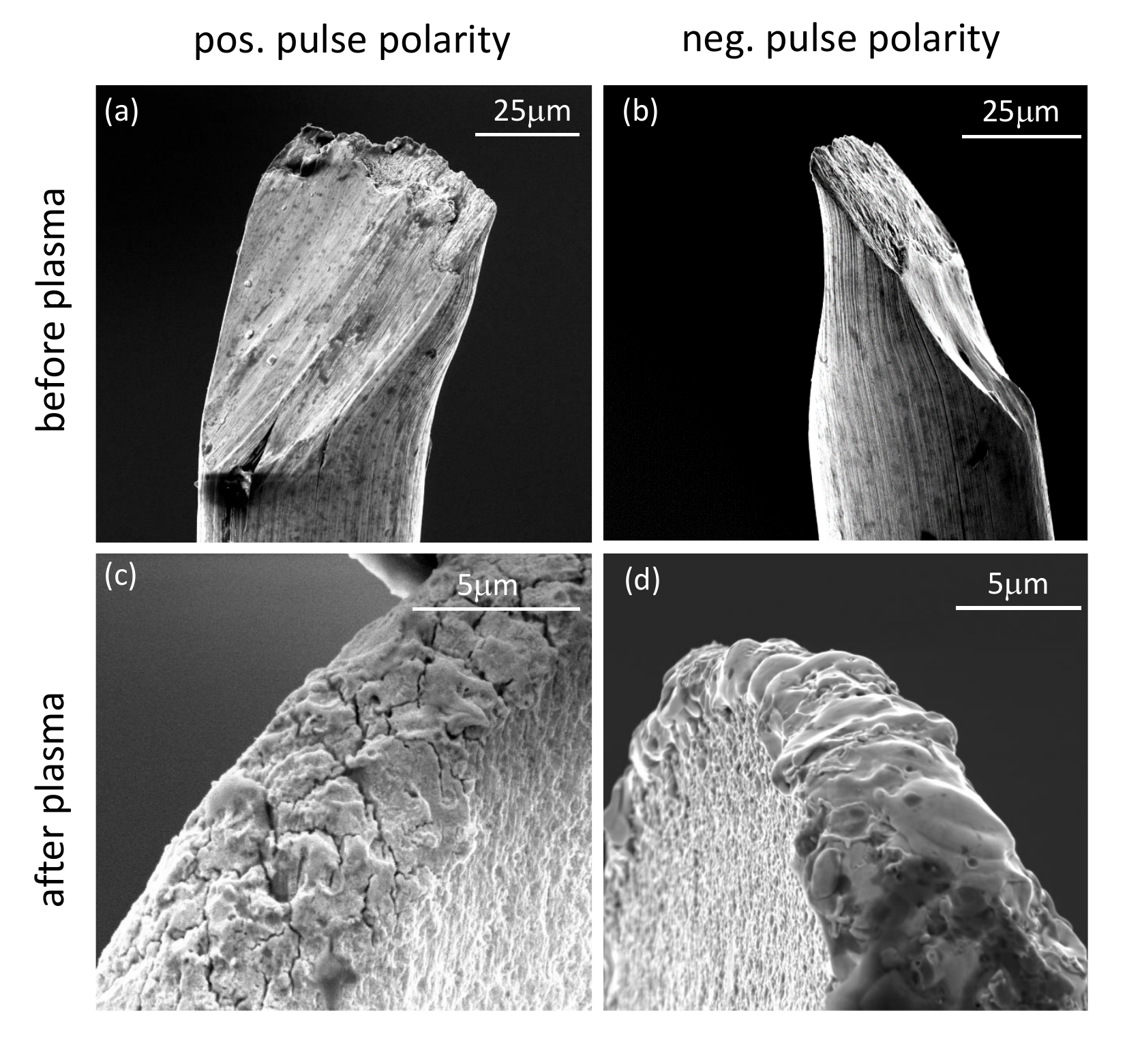}
    \caption{SEM micrographs of the tungsten electrode before (a,b) and after (c,d) plasma exposure for positive (a,c) and negative (b,d) polarity.}
    \label{fig:sem}
\end{figure}

The tungsten electrode tip surface is investigated before and after discharge treatment. Therefore, new tungsten electrode tips are taken for the measurements with positive and the negative applied pulses, respectively. The surfaces of the tungsten electrode are inspected using SEM images, as shown in Fig. \ref{fig:sem} for the positive polarity of + 20 kV before (a) and after plasma exposure for 30\,min (b) and for the negative polarity of - 20 kV before (b) and after plasma exposure of 25\,min (d). It can be seen that the degree of melting at the electrode surface is more severe in case of using the negative polarity compared to the positive polarity. 

The interpretation of the black body radiation indicates temperatures of the order of 7000 K for the positive polarity which corresponds approximately to the boiling temperature of tungsten and much lower for the negative polarity. The observation of temperatures close to that for the phase transitions of tungsten can easily be explained. Power that is dissipated during plasma ignition or propagation cause a heating of the surface up to a temperature that eventually rests at the temperature of these phase transitions, because any further power input is dissipated by completing the phase transition rather than further heating of the electrode. 

It is assumed that the electron current that is drawn by the positively biased electrode by \textit{field ionization} of water molecules, causes an intense heating which keeps the tungsten surface at the phase transition liquid-to-vapor. Here, tungsten is so efficiently evaporated that a solid surface remains without any visible melting topology at the surface after cooling down after the pulse. The high temperatures may also favor crystallization of tungsten and the creation of larger grains. The inherent stress of these grains becomes visible after cooling down as small cracks at the plasma exposed tungsten surface (see Fig. \ref{fig:sem}c).

In the case of negative polarity, however, electrons are emitted from the tungsten electrode by \textit{field emission}. This eventually cools the surface and only the Ohmic heating of the current passing the tip may heat the electrode at least up to the phase transition from solid-to-liquid. After the end of the plasma pulse, the molten tip solidifies again and molted droplets at the surface remain visible (see Fig. \ref{fig:sem}d). 

\subsection{Plasma dynamic}

\begin{figure}[ht]
    \centering
    \includegraphics[width=11cm]{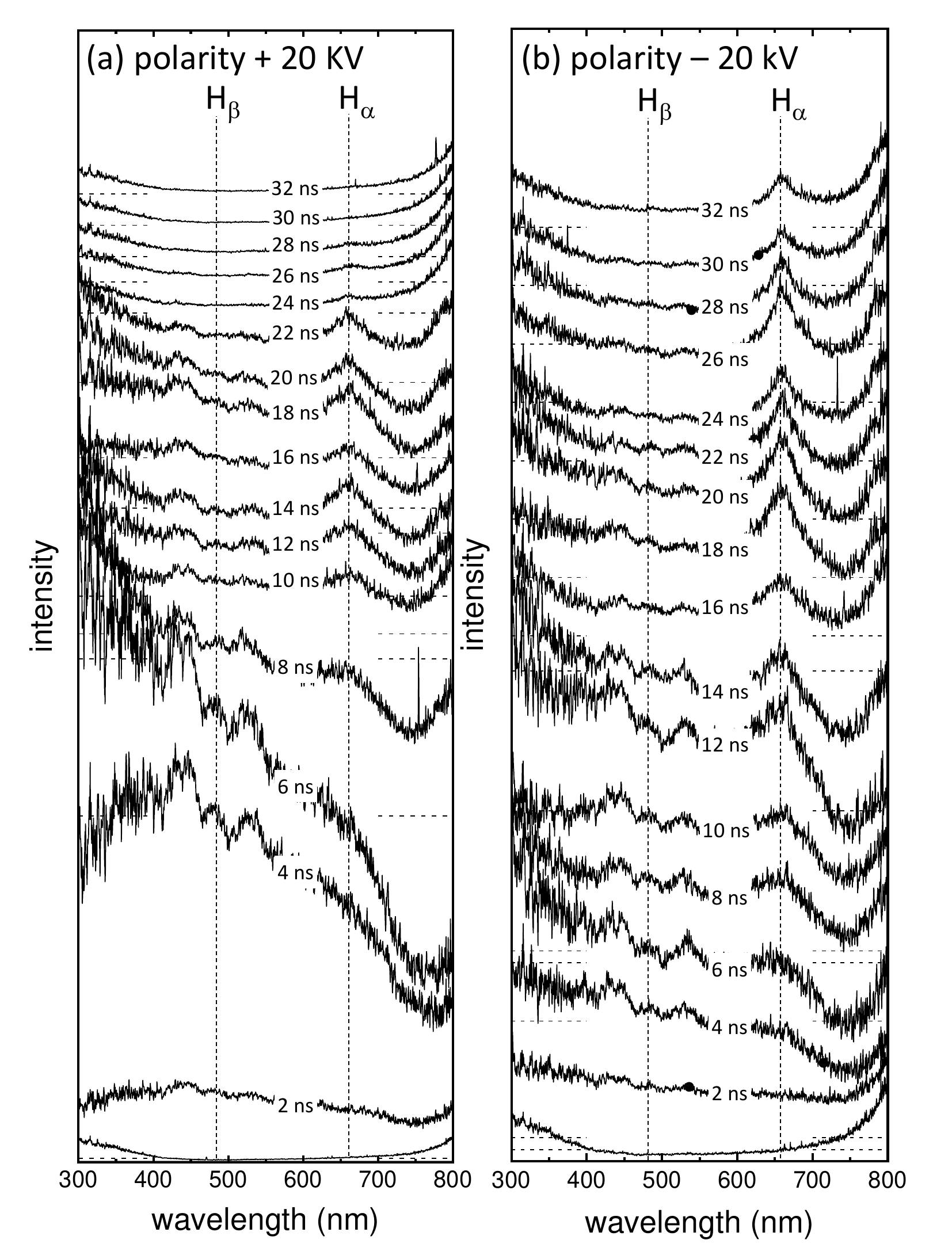}
    \caption{Temporal evolution of plasma emission for positive (a) and negative (b) polarity. The line positions for the \Ha~ and \Hb~lines are indicated. The spectra are offset for better visibility (horizontal dashed lines).}
    \label{fig:spectra}
\end{figure}

Fig. \ref{fig:spectra} shows the emission spectra for positive (a) and negative polarity (b) versus time. It can be seen that the continuum background is more pronounced for the positive polarity compared to the negative polarity, as already illustrated for the example spectra in Fig. \ref{fig:spectrum2}. The broad \Ha~emission can clearly be identified. The continuum is separated from line emission following the procedure from \cite{Grosse.2020,vonKeudell.2020}, as also illustrated before. The background consists of a black body radiation component with a temperature of 7000 K for the positive and a lower temperature for the negative polarity. A short wavelength contribution is also visible, which may be linked to the formation of a hot spot on the tungsten electrode or to Bremsstrahlung. Here, we only analyse the broadening and intensity of the \Ha~emission lines. 

\begin{figure}[ht]
    \centering
    \includegraphics[width=11cm]{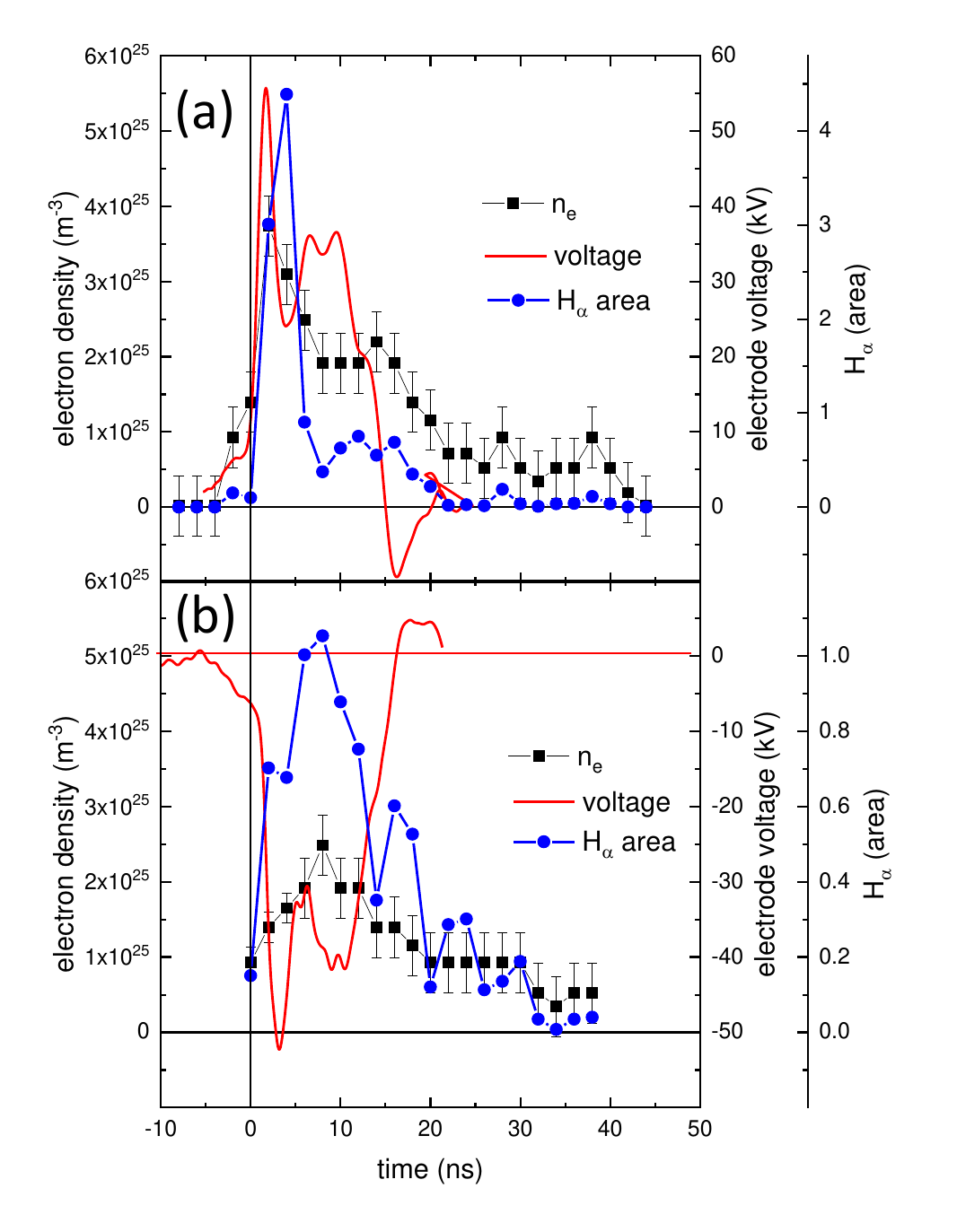}
    \caption{Temporal development of the electron densities in relation to the applied voltage for positive (a) and negative polarity (b). Please note that the scales for voltage and electron densities are identical in (a) and (b), but the scale for the \Ha~emission differs by a factor 4.}
    \label{fig:electrondensity}
\end{figure}

Fig. \ref{fig:electrondensity} shows the electron densities derived from the evaluation of the Stark broadened \Ha~lines and the number of emitting hydrogen atoms as derived from the integrated line profile of \Ha~(blue circles in Fig. \ref{fig:electrondensity}). In addition the electrode voltage is shown (red lines in Fig. \ref{fig:electrondensity}).

It can be seen that the electron density follows the electrode voltage rather closely, with electron densities reaching values up to 4 $\cdot$ 10$^{25}$ m$^{-3}$ for the positive polarity and up to 2 $\cdot$ 10$^{25}$ m$^{-3}$ for the negative polarity. After the end of the plasma pulse the electron density decays with a time constant of approximately 15 ns, which is similar to the decay of the electrical power. Especially, at times larger than 15 ns, oscillations of the voltage become visible (ringing) that eventually corresponds also to an oscillating power input causing the light emission to oscillate as well. Apparently, the electron density and its dynamic is rather similar when comparing positive and negative polarity. The line reversal of the \Ha~line is a measure for the degree of self absorption of the \Ha~light. This contribution may be visible only in the very beginning and is again rather similar for both polarities (not shown).

The number of emitting hydrogen atoms is estimated from the \Ha~line profile integral. At the very beginning, the positive polarity causes an intense light emission of \Ha, whereas it is small in case of the negative polarity. At  later stages, this integrated \Ha~light is again rather similar, when comparing the positive and negative polarity.  

\newpage
\section{Discussion}

\subsection{Different plasma characteristics for positive vs. negative polarity}

The experiments comparing negative and positive polarity applied to a tungsten electrode immersed in distilled water revealed that some characteristics are almost identical, whereas others differ significantly. The latter are associated with the interface solid-liquid and the differences between \textit{field emission} and \textit{field ionization} that cause plasma ignition. The experimental observations can be summarized as follows:

\begin{itemize}
    \item \textit{Black body temperature:} in case of a positive polarity, a continuum background equivalent to a surface temperature of 7000 K can be seen. This temperature is equivalent to the phase transition liquid-vapor of tungsten. In case of a negative polarity, a significant contribution of black body radiation cannot uniquely be identified so that the temperature should be significantly lower. Despite a similar dissipated energy, the heating of the electrode is apparently more significant for the positive polarity.
    
    \item \textit{Electron density:} the electron densities differ only by a factor of two with respect to its maximum value at the very beginning of the pulse. The densities do not depend on the polarity of the electrode for most of the later temporal development, as illustrated again in Fig. \ref{fig:necomparison}. Apparently, the efficiency of ionization during plasma propagation does not depend significantly on the direction of the electric field. This is very different to gaseous streamer discharges, where negative streamers exhibit much higher electron densities compared to positive streamers. The decay of the electron density after the HV pulse is also almost identical, as shown in Fig. \ref{fig:necomparison}. The loss of electrons may correspond to the recombination of electrons with ions indicating that the densities are again rather similar leading to an almost identical decay rate.
    
    \begin{figure}[ht]
    \centering
    \includegraphics[width=9cm]{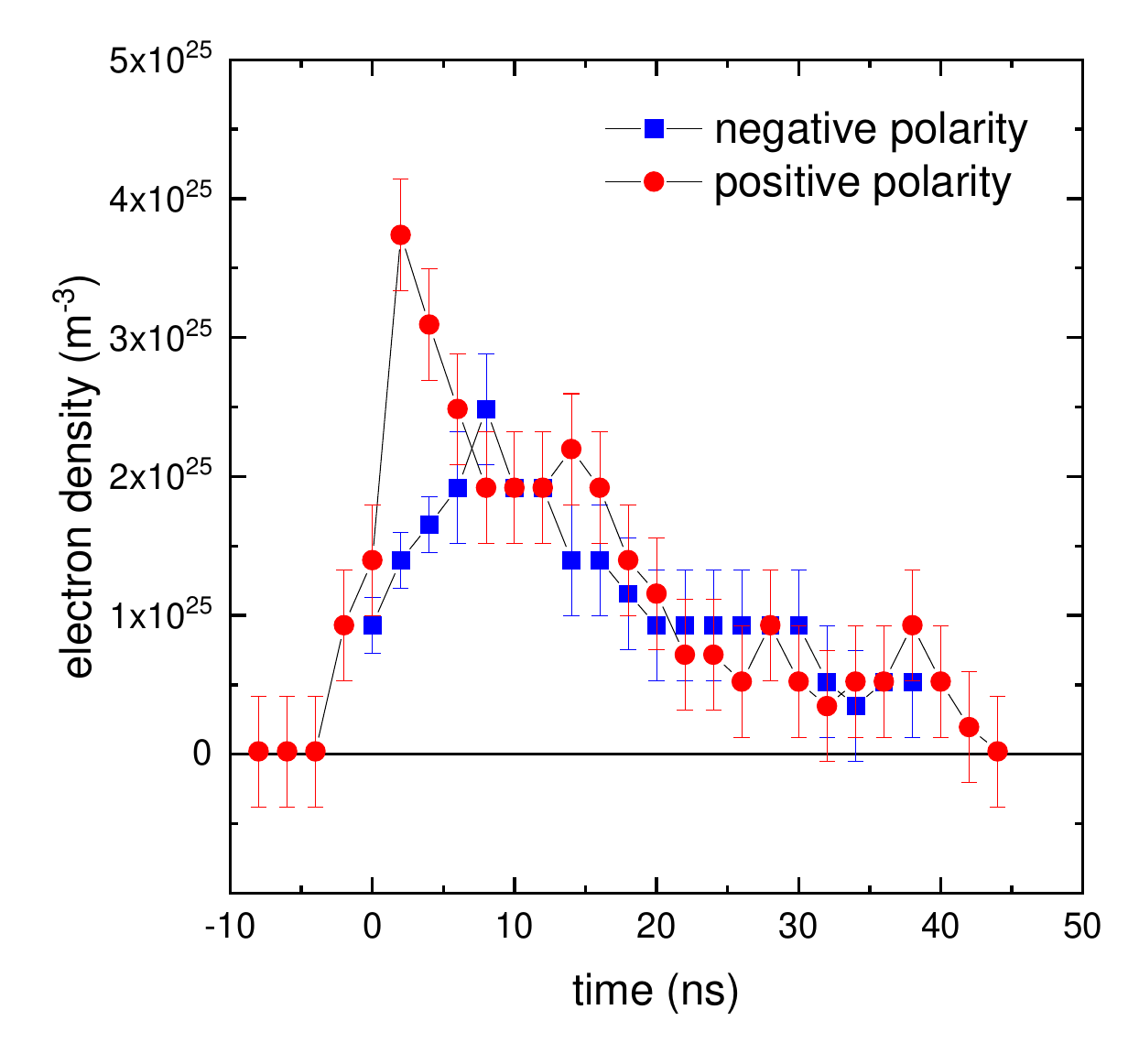}
    \caption{Comparison electron density for positive and negative polarity (same data as in Fig. \ref{fig:electrondensity})}
    \label{fig:necomparison}
    \end{figure}

    \item \textit{Emission pattern:} the emission pattern in the ICCD images are rather similar for both polarities. This is in contrast to similar experiments from Seepersad et al. \cite{Seepersad.2013a}, who reported that the emission of a positive and negative pulsed discharge inside distilled water shows very different emission patterns. The negative pulsed discharge led to a faint glowing structure at the electrode, whereas a positive pulse resulted in a filamentary discharge structure. This was connected to a gas discharge propagation. The main difference to this work is the longer rise time of 4 ns and a pin-to-plane geometry in the experiments of Seepersad et al.
    
    \item \textit{H density:} the total number of emitting hydrogen atoms is almost identical for the negative and positive polarity. Only in the very beginning, the emission by H atoms is enhanced for the positive polarity.
\end{itemize}

Based on these data, we postulate that the process of \textit{plasma ignition} at the electrode-liquid interface induces large differences in the plasma characteristics when comparing both polarities, whereas the actual \textit{plasma propagation} appears to be rather similar, as discussed in the following.

\subsection{Plasma ignition by field effects at the electrode-liquid interface}

The ignition of the plasma is caused either by \textit{field ionization of water molecules} for positive polarity or by \textit{field emission of electrons} from the electrode for negative polarity. 
    
In case of \textit{field ionization} for positive polarity, the layer of water molecules adjacent to the surface is efficiently ionized by tunneling of electrons into the electrode. In addition, free electrons from the ionization of water molecules in the high electric field are accelerated towards the tungsten electrode and cause an intense heating. At the same time, a large number of hydrogen atoms is created by for example impact dissociation or dissociative recombination of water ions and electrons. The created excited hydrogen atoms cause an intense \Ha~emission at the very beginning. At later times, when the plasma propagates along the streamer channel, the efficiency of hydrogen excitation becomes smaller.
    
In case of \textit{field emission} for negative polarity, an efficient injection of electrons into water occurs. This field emission appears to create a much smaller electron density in the beginning and also a much smaller number of hydrogen atoms that are being excited. At later times, when the plasma propagates along the streamer channel, the efficiency of hydrogen excitation is similar to that for positive polarity.
    
Apparently, \textit{field ionization} and \textit{field emission} create very different electron densities and thus also different numbers of excited hydrogen atoms during the rising front of the pulse. At the same time the heating of the tungsten electrode is much stronger for positive polarity reaching temperatures of 7000 K and only much lower temperatures for negative polarity.
    
The thresholds in electric field strengths for \textit{field ionization} and \textit{field emission} for the W/H$_2$O system are rather similar, namely 0.2 V/\AA\, \cite{Schmidt.1964, Anway.1969} and 0.3 V/\AA\, \cite{Gomer.1972}, respectively.  It is assumed that these electric fields are reached at small protrusions on the tungsten electrode. The electric field $E$ can be estimated from the applied voltage $U$ for a surface with curvature radius $r$ as $E = U/(5 r)$ \cite{Gomer.1972}. Thereby, a critical field strength of 0.2 V/\AA\, at a voltage of $U$ = 20 kV, requires a curvature radius of at least $r = 2\,\mu$m of small protrusions at the tungsten tip. Such small features can, in fact, be seen in the SEM images in Fig. \ref{fig:sem}. The threshold field strengths, however, depend also sensitively on the work function of the tungsten surface, which is 4.5 eV for elemental tungsten, but can rise to 6 eV for oxidized tungsten. Consequently, the local oxidation stage at the surface may have a significant influence on the field effects.
    
The experiments revealed smaller electron densities and a smaller black temperatures at plasma ignition for negative polarity, where \textit{field emission} dominates rather than \textit{field ionization}. We argue that this difference is associated with the difficulty of electron emission from a partly oxidized rough tungsten surface in comparison to \textit{field ionization} of an adjacent layer of water molecules. The ionization of water molecules may occur across a large surface area in front of the electrode, whereas it is assumed that field emission of electrons may occur only at few spots on the surface, where the work function is the smallest locally. As a result, the current and the plasma power is localized at a few hot spots at a negatively biased surface, which causes a smaller electron density and less heating of the tungsten electrode.
    
\subsection{Plasma ignition by volume effects}

Plasma ignition by field effects at the electrode-liquid interface may provide some seed electrons, but also volume effects may trigger plasma ignition in the liquid medium surrounding the tungsten tip. It is for example conceivable to assume electron multiplication in low density regions in the liquid with the superimposed high electric field. Such low density regions may either be nanovoids caused by liquid rupture or density fluctuation in an SCF at the critical point.

\begin{itemize}
    \item \textit{Ignition in nanovoids:} Nanovoids are being created by the electric field pressure gradient surrounding the tungsten electrode tip above a value of 2 $\cdot$ 10$^7$ Pa for  cavitation to occur to induce liquid ruptures \cite{Herbert.2006}. A seed electron may be generated at the inner walls of such a nanovoid due to field effects. Due to the very high electric fields, an extension of a few nanometer of these nanovoids is already enough to allow a successful next ionization of such an accelerated electron inside a nanovoid. Simek et al. \cite{Simek.2017} showed emission spectra for experiments using a very high voltage of 100 kV applied to a pin electrode in water, where the time dependence of the spectra evolution in the first 3 ns could be best described by electron-neutral Bremsstrahlung that develops in time. They also state that the electron generation is assumed to occur via field effects at nanovoid interfaces (or at the electrode itself) and the radiation is caused by the acceleration of these created electrons inside nanovoids. Such a region of slightly lower density is also consistent with the density estimates from the observed degree of self absorption of the hydrogen Balmer lines, as discussed previously \cite{vonKeudell.2020}. Quantitative modeling of ignition by Li et al. \cite{Li.2020} showed, however, that electron multiplication in a single nanovoid may occur, but that the density of nanovoids is only 1\% so that the development of a complete charge avalanche is difficult.
    
    \item \textit{Ignition in an SCF:} as an alternative, ignition may occur as a volume process by regarding the possible transition of the medium into a super critical state. The critical pressure of water is 2.2 $\cdot$ 10$^7$ Pa, which is almost identical to the threshold for cavitation and liquid ruptures, the critical temperatures is 647 K. Ignition as a volume process my follow three steps in time: (i) at first, local field effects at the electrode-liquid interface cause ignition that induces a high local pressure of typically 10$^9$ Pa (estimated from the boundary condition of cavitation modelling above), but also a temperature of the local medium above the critical point of water (estimated from the black body radiation). (ii) Second, a pressure and temperature field expands into the medium surrounding the electrode tip, as illustrated in Fig. \ref{fig:pressurefield} showing a simple $r^{-1}$ dependence \cite{GilmoreF.R..1952} of the pressure surrounding a bubble with radius of 25 $\mu$m. The interface between the high pressure field and the ambient liquid propagates with the sound velocity outwards after ignition. The velocity of the pressure wave propagating into the liquid within the first 44 ns after ignition had been estimated from the shadowgraphy measurements and ranges between 5500 ms$^{-1}$ to 9000 ms$^{-1}$ depending on the electrode voltage \cite{Grosse.2019}. This can be converted into a region with a radius of 55 $\mu$m and 90 $\mu$m that is affected within 10 ns and that can be converted into the super critical state. (iii) Third, at the front of the acoustic wave, the pressure goes from a value above the critical pressure (marked as red line at 2.2 $\cdot$ 10$^7$ Pa in the Fig. \ref{fig:pressurefield}) to the ambient pressure. Therefore, in the region of high electric field an acoustic wave front travels outwards that induces the liquid to go through the critical pressure. Finally, the density fluctuations at the critical point at this acoustic wave front allow ignition. This may be compared with a shadowgraphic image of the discharge taken at 44 ns after ignition but with a gate time of 70 ns to cover the expansion of the pressure field but also plasma emission, as shown in the insert in Fig. \ref{fig:pressurefield}. One can clearly see a dark area surrounding the tip as an indication of the high pressure field and also that plasma propagation itself occurs  \textit{inside} this region. It can be assumed that ignition occurs within such a region. A similar connection between the region affected by the pressure and the region where the first streamers becomes visible \cite{Seepersad.2013}.
 
     \begin{figure}[ht]
    \centering
    \includegraphics[width=11cm]{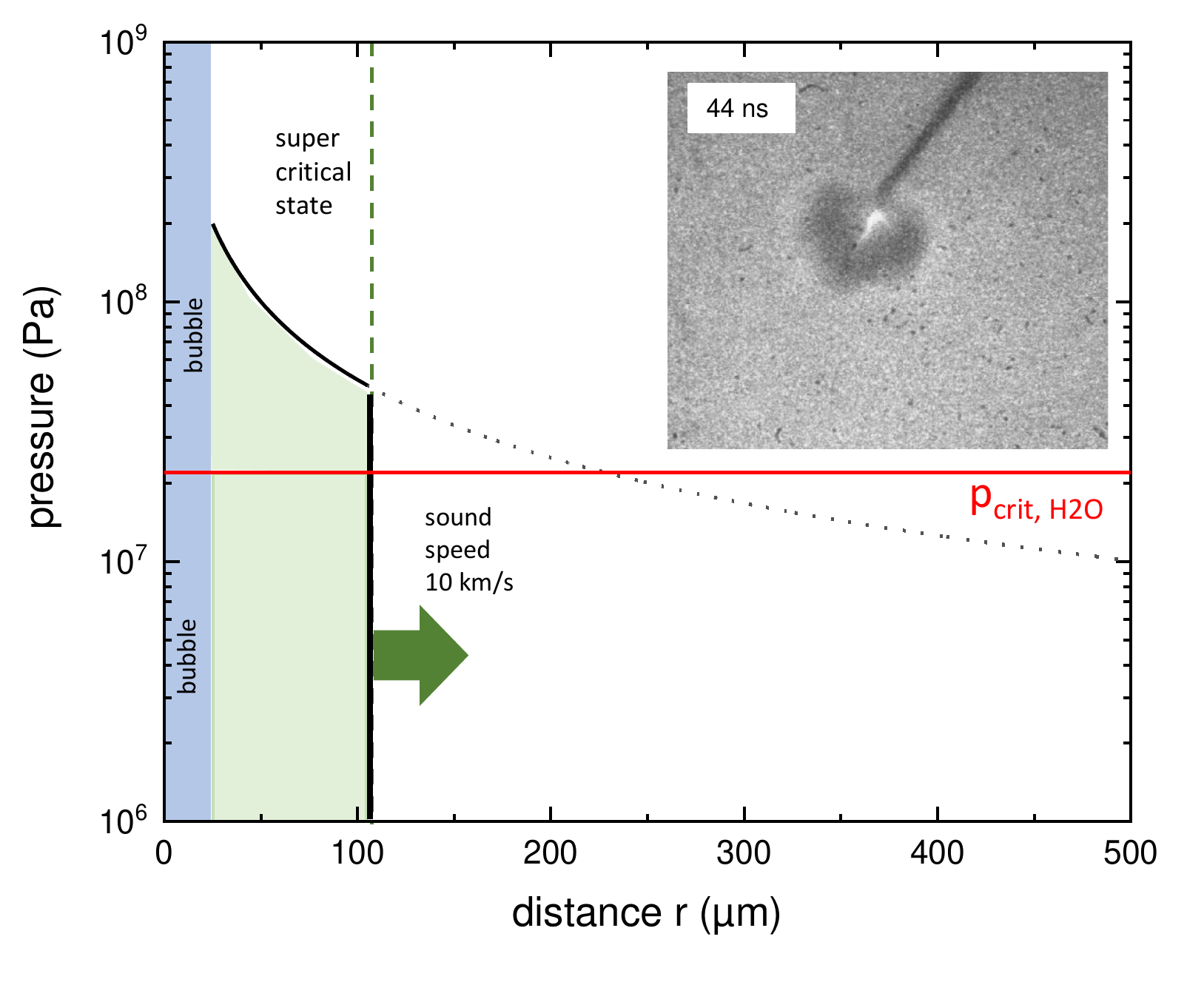}
    \caption{Pressure field in the liquid after ignition according to cavitation theory. The inserts show shadowgraphy of the ignition phase with a delay of 44\,ns and a camera gate of 70\,ns. The red line denotes the critical pressure for water.}
    \label{fig:pressurefield}
    \end{figure}  
 
    \end{itemize}
    
The ignition via nanovoids or via a SCF should occur rather similarly. However, the experiments reveal large differences with respect to the plasma characteristics upon ignition. Therefore, we argue that field effects at the interface-liquid should be dominating.

\subsection{Plasma propagation}

Plasma propagation for positive and negative polarity are almost identical with respect to the electron density or the number of exited hydrogen atoms, as a measure for the density of the medium in which the streamers propagate. Based on the impact of self absorption on line emission, we assume that the density is slightly lower than the liquid density \cite{vonKeudell.2020}. Any independence of plasma propagation on the polarity is consistent with our hypothesis of an ionization mechanism in model B and D consisting of electron tunneling in between adjacent water molecules or electron channeling in an SCF, respectively. To some extent also model C may be still consistent with the data, because the acceleration of electrons in small nanovoids may also not be affected by the direction of the electric field at least in first order. Any gradient in the density of the nanovoids may exist, but since the data for the plasma propagation are almost identical, any influence of such a gradient cannot be seen. Nevertheless, the presented data do not allow to distinguish between plasma propagation via field effects or via acceleration of electrons in an ensemble of nanovoids.
      
The identical electron densities can be explained by assuming that the local ionization processes are the same for both polarities. Nevertheless, the data on the \Ha~emission intensity show that the light in case of the negative polarity takes a bit longer time to decay after the pulse in comparison to the positive polarity. One may speculate that this is caused by different volumina that are being affected by the ionization fronts propagating in different direction for both polarities either away from or towards the electrode tip, as discussed for model D, although more detailed experiment and modeling is required to be able to interpret this.

\subsection{Model for ignition and plasma propagation}

The hypothesis for plasma ignition and propagation is most consistent with model D and is summarized in the following 
. The plasmas are ignited via field ionization of water molecules at the electrode-liquid interface for positive polarity. This ionization occurs over a large area in front of the electrode surface. The collected electrons cause a strong heating of the tungsten tip. For negative polarity, ignition occurs via field emission which is more localized at the electrode surface. The initial pressures are of the order of 10$^9$ Pa and temperatures in the range of 1000 of K. This converts the adjacent water into the super critical state. In such a state, electron acceleration and multiplication may occur via electron channeling in these density fluctuations. One may coin such plasma propagation as SCF-streamers that propagate towards the tungsten tip for positive and away from the tip for negative polarity. The electron densities do not depend on the direction of propagation of the SCF-streamers. This interpretation is consistent with the data, but remain still be an indirect reasoning and more detailed time and space resolved experiments as well as multi physics modeling on an atomistic scale will be required to uniquely identify the plasma ignition and propagation mechanisms in the future. 

\section{Conclusion}

Plasma ignition and plasma propagation for nanosecond plasmas in water have been analysed by optical and electrical diagnostics comparing positive and negative polarity of $\pm$ 20 kV applied to a tungsten electrode. It is shown that the plasma ignition by field ionization creates a larger electron density and thus a larger density of exited hydrogen atoms compared to plasma initiation by field emission. The heating of the tungsten electrode is much more severe for positive polarity reaching temperatures up to 7000 K, whereas the temperatures are much lower for negative polarity. The electron densities associated with plasma propagation, however, are rather similar. This led to the conclusion that plasma propagation is a very local effect for plasmas inside liquids that is either governed by field ionization of water molecules around the streamer head or by acceleration and field emission of electrons in nanovoids or sustained by density fluctuations in water due to its super critical state.
\newpage

\section*{Acknowledgements}
The authors appreciate the help from Susanne Jordans and Felicitas Scholz from Chair of Materials Science and Engineering, Ruhr-University Bochum for their assistance with the SEM images.
This project is supported by the DFG (German Science Foundation) within the framework of the Coordinated Research Centre SFB 1316 at Ruhr-University Bochum. 


\section*{References}

\bibliography{PlasmaLiquids} 

\providecommand{\newblock}{}
\begin{thebibliography}{10}
\expandafter\ifx\csname url\endcsname\relax
  \def\url#1{{\tt #1}}\fi
\expandafter\ifx\csname urlprefix\endcsname\relax\def\urlprefix{URL }\fi
\providecommand{\eprint}[2][]{\url{#2}}

\bibitem{Bruggeman.2009}
Bruggeman P~J and Leys C 2009 {\em Journal of Physics D: Applied Physics\/}
  {\bf 42} 053001 ISSN 0022-3727

\bibitem{Seepersad.2013}
Seepersad Y, Pekker M, Shneider M~N, Fridman A~A and Dobrynin D 2013 {\em
  Journal of Physics D: Applied Physics\/} {\bf 46} 355201 ISSN 0022-3727

\bibitem{Joshi.2009}
Joshi R~P, Kolb J~F, Xiao S, Schoenbach K~H and Schoenbach K~H 2009 {\em Plasma
  Processes and Polymers\/} {\bf 6} 763--777 ISSN 16128850

\bibitem{Zhuang.2016}
Zhuang J, Sun A and Huo C 2016 {\em High Voltage\/} {\bf 1} 74--80 ISSN
  2397-7264

\bibitem{Sharbaugh.1978}
Sharbaugh A, Devins J and Rzad S 1978 {\em IEEE Transactions on Electrical
  Insulation\/} {\bf EI-13} 249--276 ISSN 0018-9367

\bibitem{Simek.2020}
{\v S}imek M, Hoffer P, Tungli J, Prukner V, Schmidt J, B{\'i}lek P and
  Bonaventura Z 2020 {\em Plasma Sources Science and Technology\/} {\bf 29}
  064001 ISSN 1361-6595

\bibitem{Hoffer.2020}
Hoffer P, Prukner V, Schmidt J and {\v S}imek M 2020 {\em Japanese Journal of
  Applied Physics\/} {\bf 59} SHHA08 ISSN 0021-4922, 1347-4065

\bibitem{Devins.1981}
Devins J~C, Rzad S~J and Schwabe R~J 1981 {\em Journal of Applied Physics\/}
  {\bf 52} 4531--4545 ISSN 0021-8979, 1089-7550

\bibitem{OSullivan.2008}
O'Sullivan F, Hwang J, Zahn M, Hjortstam O, Pettersson L, {Rongsheng Liu} and
  Biller P 2008 A {{Model}} for the {{Initiation}} and {{Propagation}} of
  {{Positive Streamers}} in {{Transformer Oil}} {\em Conference {{Record}} of
  the 2008 {{IEEE International Symposium}} on {{Electrical Insulation}}\/}
  ({Vancouver, BC}: {IEEE}) pp 210--214 ISBN 978-1-4244-2091-9

\bibitem{Grosse.2020}
Grosse K, {Schulz-von der Gathen} V and {von Keudell} A 2020 {\em Plasma
  Sources Science and Technology\/} {\bf 29} 095008 ISSN 1361-6595

\bibitem{vonKeudell.2020}
{von Keudell} A, Grosse K and {Schulz-von der Gathen} V 2020 {\em Plasma
  Sources Science and Technology\/} {\bf 29} 085021 ISSN 1361-6595

\bibitem{Nijdam.2020}
Nijdam S, Teunissen J and Ebert U 2020 {\em Plasma Sources Science and
  Technology\/} {\bf 29} 103001 ISSN 1361-6595

\bibitem{Fridman.2005}
Fridman A, Chirokov A and Gutsol A 2005 {\em Journal of Physics D: Applied
  Physics\/} {\bf 38} R1--R24 ISSN 0022-3727, 1361-6463

\bibitem{Tereshonok.2017}
Tereshonok D~V 2017 {\em Journal of Physics D: Applied Physics\/} {\bf 50}
  015603 ISSN 0022-3727, 1361-6463

\bibitem{Tereshonok.2018}
Tereshonok D~V, Babaeva N~Y, Naidis G~V, Panov V~A, Smirnov B~M and Son E~E
  2018 {\em Plasma Sources Science and Technology\/} {\bf 27} 045005 ISSN
  1361-6595

\bibitem{Schmidt.1964}
Schmidt W~A 1964 {\em Zeitschrift f\"ur Naturforschung A\/} {\bf 19} 318--327
  ISSN 1865-7109, 0932-0784

\bibitem{Anway.1969}
Anway A~R 1969 {\em The Journal of Chemical Physics\/} {\bf 50} 2012--2021 ISSN
  0021-9606, 1089-7690

\bibitem{Shneider.2012}
Shneider M, Pekker M and Fridman A 2012 {\em IEEE Transactions on Dielectrics
  and Electrical Insulation\/} {\bf 19} 1579--1582 ISSN 1070-9878

\bibitem{Shneider.2013}
Shneider M~N and Pekker M 2013 {\em Physical Review E\/} {\bf 87} ISSN
  1539-3755, 1550-2376

\bibitem{Li.2020}
Li Y, Li L, Wen J, Zhang J, Wang L and Zhang G 2020 {\em Plasma Sources Science
  and Technology\/} {\bf 29} 075005 ISSN 1361-6595

\bibitem{Aghdam.2020}
Aghdam A~C and Farouk T 2020 {\em Plasma Sources Science and Technology\/} {\bf
  29} 025011 ISSN 1361-6595

\bibitem{Stauss.2018}
Stauss S, Muneoka H and Terashima K 2018 {\em Plasma Sources Science and
  Technology\/} {\bf 27} 023003 ISSN 1361-6595

\bibitem{Grosse.2019}
Grosse K, Held J, Kai M and {von Keudell} A 2019 {\em Plasma Sources Science
  and Technology\/} {\bf 28} 085003 ISSN 1361-6595

\bibitem{Muneoka.2015}
Muneoka H, Urabe K, Stauss S and Terashima K 2015 {\em Physical Review E\/}
  {\bf 91} 042316 ISSN 1539-3755, 1550-2376

\bibitem{Ito.2002}
Ito T and Terashima K 2002 {\em Appl. Phys. Lett.\/} {\bf 80} 2854

\bibitem{Pongrac.2018}
Pongr{\'a}c B, {\v S}imek M, {\v C}lupek M, Babick{\'y} V and Luke{\v s} P 2018
  {\em Journal of Physics D: Applied Physics\/} {\bf 51} 124001 ISSN 0022-3727,
  1361-6463

\bibitem{Marinov.2014}
Marinov I, Starikovskaia S and Rousseau A 2014 {\em Journal of Physics D:
  Applied Physics\/} {\bf 47} 224017 ISSN 0022-3727

\bibitem{Simek.2017}
{\v S}imek M, Pongr{\'a}c B, Babick{\'y} V, {\v C}lupek M and Luke{\v s} P 2017
  {\em Plasma Sources Science and Technology\/} {\bf 26} 07LT01 ISSN 1361-6595

\bibitem{GilmoreF.R..1952}
Gilmore F 1952 The growth or collapse of a spherical bubble in a viscous
  compressible liquid Tech. Rep. 26-4 {Hydrodynamic Laboratory, Caltech}
  {Pasadena, California}

\bibitem{Plesset.1949}
Plesset 1949 {\em J Applied Mechanics\/}  277

\bibitem{Plesset.1977}
Plesset M~S and Prosperetti A 1977 {\em Ann. Rev. Fluid Mech.\/} {\bf 9} 145

\bibitem{Keller.1980}
Keller J~B and Miksis M 1980 {\em J. Acoust. Soc. Am.\/} {\bf 66} 628

\bibitem{Keller.1956}
Keller J~B and Kolodner I~I 1956 {\em Journal of Applied Physics\/} {\bf 27}
  1152--1161 ISSN 0021-8979, 1089-7550

\bibitem{Gigosos.2003}
Gigosos M~A, Gonz{\'a}lez M~{\'A} and Carde{\~n}oso V 2003 {\em Spectrochimica
  Acta Part B: Atomic Spectroscopy\/} {\bf 58} 1489--1504 ISSN 0584-8547

\bibitem{Cowan.1948}
Cowan R~D and Dieke G~H 1948 {\em Reviews of Modern Physics\/} {\bf 20}
  418--455 ISSN 0034-6861

\bibitem{Seepersad.2013a}
Seepersad Y, Pekker M, Shneider M~N, Dobrynin D and Fridman A 2013 {\em Journal
  of Physics D: Applied Physics\/} {\bf 46} 162001 ISSN 0022-3727, 1361-6463

\bibitem{Gomer.1972}
Gomer R 1972 {\em Accounts of Chemical Research\/} {\bf 5} 41--48 ISSN
  0001-4842, 1520-4898

\bibitem{Herbert.2006}
Herbert E, Balibar S and Caupin F 2006 {\em Physical Review E\/} {\bf 74} ISSN
  1539-3755, 1550-2376

\end{thebibliography}

\end{document}